\documentclass[12pt,draftclsnofoot,onecolumn]{IEEEtran}
\IEEEoverridecommandlockouts
\usepackage{booktabs}
\usepackage{enumerate}
\makeatletter
\newsavebox{\@tabnotebox}

\makeatother	

\newif\ifhavebib

\usepackage{blindtext}
\usepackage{graphicx}
\usepackage{graphicx}
\usepackage{array}
\usepackage{url}
\usepackage{algorithmic}
\usepackage[cmex10]{amsmath}
\usepackage{ifpdf}
\usepackage{amssymb,amsmath}
\usepackage{empheq}
\usepackage{stfloats}   
\usepackage{textcomp}
\usepackage{cite}
\usepackage{multirow}
\usepackage{diagbox}
\usepackage{subcaption}
\let\oldFootnote\footnote
\newcommand\nextToken\relax

\renewcommand\footnote[1]{%
    \oldFootnote{#1}\futurelet\nextToken\isFootnote}

\newcommand\isFootnote{%
    \ifx\footnote\nextToken\textsuperscript{,}\fi}

\usepackage[ruled,vlined,lined,ruled,boxed]{algorithm2e}
\usepackage[dvipsnames,usenames]{color}
\usepackage[normalem]{ulem}
\usepackage{amsthm}

\usepackage{graphicx}
\usepackage{ifpdf}
\usepackage{cite}
\usepackage{enumerate}
\usepackage{enumitem}

\usepackage{array}
\usepackage{mdwmath}
\usepackage{mdwtab}
\usepackage{eqparbox}
\usepackage{bm}

\usepackage{setspace}
\usepackage{booktabs}
\usepackage[subnum]{cases}
\usepackage{rotating}

\usepackage{listings} 
\definecolor{Red}{rgb}{1,0,0}
\definecolor{Blue}{rgb}{0,0,1}
\definecolor{Olive}{rgb}{0.41,0.55,0.13}
\definecolor{Green}{rgb}{0,1,0}
\definecolor{MGreen}{rgb}{0,0.8,0}
\definecolor{DGreen}{rgb}{0,0.55,0}
\definecolor{Yellow}{rgb}{1,1,0}
\definecolor{Cyan}{rgb}{0,1,1}
\definecolor{Magenta}{rgb}{1,0,1}
\definecolor{Orange}{rgb}{1,.5,0}
\definecolor{Violet}{rgb}{.5,0,.5}
\definecolor{Purple}{rgb}{.75,0,.25}
\definecolor{Brown}{rgb}{.75,.5,.25}
\definecolor{Grey}{rgb}{.5,.5,.5}

\newcommand{\boxhead}[5]{
   \pagestyle{myheadings}
   \thispagestyle{plain}
   \setcounter{page}{1}
   \noindent
   \begin{center}
   \framebox{
      \vbox{\vspace{2mm}
    \hbox to 6.28in { {\bf #1 \hfill} }
       \vspace{6mm}
       \hbox to 6.28in { {\Large \hfill \bf #2  \hfill} }
       \vspace{6mm}
       \hbox to 6.28in { {\it #3 #4 \hfill  #5} }
      \vspace{2mm}}
   }
   \end{center}
   \markboth{#5 -- #2}{#5 -- #2}
   \vspace*{4mm}
}

\usepackage[colorlinks=true, urlcolor=black,  linkcolor=black, citecolor=black]{hyperref}
\theoremstyle{definition}
\newtheorem{theorem}{Theorem} 
 
\newtheorem{corollary}{Corollary}

\newtheorem{lemma}{Lemma}

\theoremstyle{remark}
\newtheorem{remark}{Remark}

\theoremstyle{definition}

\DeclarePairedDelimiterX{\infdivx}[2]{(}{)}{%
	#1\;\delimsize\|\;#2%
}

\DeclarePairedDelimiter{\norm}{\lVert}{\rVert}
\DeclarePairedDelimiter{\abs}{\lvert}{\rvert}

\DeclareMathOperator*{\argmin}{\mathop{\arg\min}}
\def\diag{\mathop{\rm diag}\nolimits}%
\def\Re{\mathop{\rm Re}\nolimits}%
\newcommand{\av}{{\bf a}}
\newcommand{\bv}{{\bf b}}
\newcommand{\ev}{{\bf e}}

\newcommand{\Xv}{{\bf X}}

\newcommand{\rv}{{\bf r}}

\newcommand{\Hv}{{\bf H}}
\newcommand{\Gv}{{\bf G}}

\newcommand{\Iv}{{\bf I}}
\newcommand{\fv}{{\bf f}}
\newcommand{\gv}{{\bf g}}
\newcommand{\xv}{{\bf x}}

\newcommand{\yv}{{\bf y}}

\newcommand{\hv}{{\bf h}}

\newcommand{\nv}{{\bf n}}

\newcommand{\zetav}{\boldsymbol \zeta}

\newcommand{\wv}{{\bf w}}

\newcommand{\thetav}{\boldsymbol \theta}

\def\a{\alpha}
\def\b{\beta}

\def\e{\epsilon}

\DeclareMathOperator\E{E}

 \def\E{\mathbb{E}}

\def\de \mathrm{d}

\newcommand{\CN}{\mathcal{CN}}

\newcommand\eg{e.g.,\xspace}
\newcommand\ie{i.e.,\xspace}
\def\textiid{i.i.d.\@\xspace}

\newcommand\iid{\ifmmode\text{ i.i.d. } \else \textiid \fi}

\newcommand{\Complex}{\mathbb{C}}
\newcommand{\Real}{\mathbb{R}}

\newcommand{\beqs}{\begin{equation*}}
\newcommand{\eeqs}{\end{equation*}}
\newcommand{\beq}{\begin{equation}}
\newcommand{\eeq}{\end{equation}}
\begin{document}



\setitemize{listparindent=\parindent,partopsep=0pt,topsep=-0.25ex}
\setenumerate{fullwidth,itemindent=\parindent,listparindent=\parindent,itemsep=0ex,partopsep=0pt,parsep=0ex}

\havebibtrue
\title{Reconfigurable Intelligent Surface Enabled Federated Learning: A Unified Communication-Learning Design Approach}
\author{
	Hang~Liu,~\IEEEmembership{Graduate Student Member,~IEEE,} Xiaojun~Yuan,~\IEEEmembership{Senior Member,~IEEE,}
	and~Ying-Jun~Angela~Zhang,~\IEEEmembership{Fellow,~IEEE}
	\thanks{H. Liu and Y.-J. A. Zhang are with the Department of Information Engineering, The Chinese University of Hong Kong, Shatin, New Territories, Hong Kong (e-mail: lh117@ie.cuhk.edu.hk; yjzhang@ie.cuhk.edu.hk). X. Yuan is with the Center for Intelligent Networking and Communications, the University of Electronic Science and Technology of China, Chengdu, China (e-mail: xjyuan@uestc.edu.cn).}
}
\maketitle
\begin{abstract}
To exploit massive amounts of data generated at mobile edge networks, \emph{federated learning} (FL) has been proposed as an attractive substitute for centralized machine learning (ML). By collaboratively training a shared learning model at edge devices, FL avoids direct data transmission and thus overcomes high communication latency and privacy issues as compared to centralized ML. To improve the communication efficiency in FL model aggregation, \emph{over-the-air computation} has been introduced to support a large number of simultaneous local model uploading by exploiting the inherent superposition property of wireless channels. However, due to the heterogeneity of communication capacities among edge devices, over-the-air FL suffers from the straggler issue in which the device with the weakest channel acts as a bottleneck of the model aggregation performance. This issue can be alleviated by device selection to some extent, but the latter still suffers from a tradeoff between data exploitation and model communication. In this paper, we leverage the \emph{reconfigurable intelligent surface} (RIS) technology to relieve the straggler issue in over-the-air FL. Specifically, we develop a learning analysis framework to quantitatively characterize the impact of device selection and model aggregation error on the convergence of over-the-air FL. Then, we formulate a unified communication-learning optimization problem to jointly optimize device selection, over-the-air transceiver design, and RIS configuration. Numerical experiments show that the proposed design achieves substantial learning accuracy improvement compared with the state-of-the-art approaches, especially when channel conditions vary dramatically across edge devices.
\end{abstract}

\section{Introduction}
The availability of massive amounts of data at mobile edge devices has led to a surge of interest in developing artificial intelligence (AI) services, such as image recognition \cite{RESNET}  and natural language processing \cite{NLP}, at the edge of wireless networks. 
Conventional machine learning (ML) requires a data center to collect all data for centralized model training. In a wireless system, collecting data from distributed mobile devices incurs
huge energy/bandwidth cost, high time delay, and potential privacy issues \cite{8970161}. To address these challenges, a new paradigm called federated learning (FL) has emerged \cite{BGD}. In a typical FL framework, each edge device computes its local model updates based on its own dataset and uploads the model updates to a parameter server (PS).  The global model is  computed at the PS and 
shared with the devices. By doing so, direct data transmission is replaced by model parameter uploading. This significantly relieves the communication burden and prevents revealing local data to the other devices and the PS.

Despite the above advantages of FL, uplink communication overhead incurred during the iterative model update process is still a critical bottleneck for FL training \cite{BGD,FEDSGD}.
To improve the communication efficiency in FL,
over-the-air computation \cite{4305404} has emerged to support analog model uploading  from massive edge devices \cite{GZhu_BroadbandAircomp,FL_1, FL_DG,FL_Digital1,FL_Digital2,zhu2020onebit}. In over-the-air model aggregation, edge devices concurrently transmit their local model updates using the same radio resources. The PS then computes the model aggregation from the received signal by exploiting the signal-superposition property of multiple-access channels. Compared with traditional orthogonal multiple access (OMA) protocols where multiple devices transmit using orthogonal channels, the bandwidth requirement or the communication latency of over-the-air computation does not increase with the number of devices, which largely relieves the communication bottleneck in FL. The first over-the-air model aggregation scheme appeared in \cite{GZhu_BroadbandAircomp}, where the authors show that the over-the-air computation leads to substantial latency reduction compared with the OMA schemes. The method in \cite{GZhu_BroadbandAircomp} was later extended to one-bit over-the-air computation in \cite{zhu2020onebit}. Moreover, the authors in \cite{FL_1} maximized the number of participants in over-the-air model aggregation subject to a specific communication error constraint. The authors in \cite{FL_DG} investigated over-the-air computation error minimization in multiple-input multiple-output (MIMO) systems. The authors in  \cite{FL_Digital1,FL_Digital2} studied over-the-air model aggregation with gradient sparsification and compression in Gaussian channels and wireless fading channels.

{Although over-the-air computation is envisioned to be a scalable FL model aggregation solution, it suffers from the straggler issue. That is, the devices with weak channels (\ie the stragglers\footnote{In the Computer Science literature, the word ``stragglers" usually refer to devices with low \emph{computation} capacities. Since the \emph{communication} aspect of FL is studied in this paper, we here use ``stragglers" to denote edge devices with poor channel conditions.})
dominate the overall model aggregation error since the devices with better channel qualities have to lower their transmit power to align the local models at the PS. 
Ref. \cite{GZhu_BroadbandAircomp} proposed to exclude the stragglers from model aggregation to avoid large communication error. Nevertheless, excluding  devices reduces the training data size, which inevitably jeopardizes  the convergence of FL.
To address this tradeoff, we must understand the impact of both the communication and computation aspects on the FL performance. However, analyzing the impact of device selection is difficult, letting alone the combined impact of device selection and communication error \cite{Friedlander2012}. The existing work adopts heuristic metrics to simplify the over-the-air FL system design. For example, Ref. \cite{GZhu_BroadbandAircomp} approximates the learning performance by the fraction of exploited data and schedules the devices with strong and weak channels to participate alternately. In \cite{FL_1},  the number of selected devices is used as a rough proxy of the FL performance and the communication-learning tradeoff is balanced by a manually tuned communication error constraint. These ad hoc solutions may not be able to fully characterize the FL training performance, leading to sub-optimal system designs.}

Reconfigurable intelligent surface (RIS), as a key enabler of the next-generation wireless networks, can reduce energy consumption and improve spectral efficiency of wireless networks \cite{wu2020intelligent,9136592,yuan2020reconfigurableintelligentsurface}. Specifically, RISs are thin sheets comprising a large number of passive reflecting elements. By inducing independent phase shifts on the incident signals, the reflecting elements can proactively manipulate the propagation channels to overcome the unfavorable propagation conditions. {Much research in recent years has focused on designing RISs to enhance traditional wireless communications. To name a few, the authors in \cite{CHUANGRIS,LIS_RZhang_discrete,9110869,ris_ref1_work4,ris_ref2_work4,ris_ref3_work4,RISMEC} have investigated the optimization of RISs in various communication systems.  As the objective of communication for over-the-air FL is different from that in traditional communication systems, there remains a need for efficient design for RIS-assisted over-the-air FL.}
The recent work in \cite{FL_RIS2} shows that the RIS can significantly mitigate the communication error in a general over-the-air computation task. Ref. \cite{FL_RIS1} made a first attempt in applying the RIS-assisted over-the-air computation method in \cite{FL_RIS2} to FL and observes considerable learning performance improvement compared with FL without RISs. 

Although the current work in \cite{FL_RIS2,FL_RIS1} has demonstrated the effectiveness of RISs on improving the over-the-air model aggregation quality, the state-of-the-art design only focuses on the communication aspect, and thus cannot fully unleash the benefits of RISs in FL systems. {Due to the aforementioned communication-learning tradeoff,  it is necessary to jointly optimize RIS configurations and device selection in a uniform framework that can directly characterize the FL performance.}

In this paper, we explore the advances of RISs in enhancing the over-the-air FL. In contrast to the existing work that considers a communication objective \cite{FL_RIS1,FL_RIS2,FL_DG}  or a rough proxy of the learning performance (\eg the number of participants as in \cite{FL_1}),  we develop a unified framework  that tracks the impact of device selection and communication error on the FL training loss under mild assumptions on the learning model. 
The contributions of this paper are summarized as follows.
\begin{itemize}
	\item We study a RIS-enabled FL system where a RIS is employed to assist the over-the-air model aggregation. We describe the optimal transmitter design in the considered system and derive a closed-form expression of the model aggregation error. 
		 
	\item We derive an upper bound on the iterative learning loss by taking errors from device selection and model aggregation into account. Based on the analysis, we show that the device selection loss and the communication error cause convergence rate reduction and lead to a non-diminishing gap between the trained model and the global optimum that minimizes the training loss. We then derive a closed-form expression of this gap, based on which we formulate a unified communication-learning optimization problem. Besides, we quantitatively analyze the communication-learning tradeoff and demonstrate the importance of RIS phase shifts in the proposed system.
	\item We propose an effective algorithm to jointly optimize device selection, receiver beamforming, and RIS phase shifts. Specifically, we employ Gibbs sampling to select the active devices and use the successive convex approximation (SCA) principle \cite{SCA} to jointly optimize the receiver beamforming and the RIS phase shifts. 
\end{itemize}
Simulation results confirm that our proposed approach achieves substantial performance improvement compared with the existing FL solutions. In particular, the proposed algorithm achieves an accuracy very close to that of the error-free ideal benchmark even when channel conditions vary significantly across devices. Besides,  our algorithm requires a much smaller RIS than the state-of-art method to achieve the same learning accuracy.

The remainder of this paper is organized as follows. In Section \ref{sec2}, we describe the FL model, the RIS-assisted communication system model, and the over-the-air model aggregation framework. In Section \ref{sec3}, we analyze the FL performance and accordingly formulate the learning optimization problem that minimizes the FL training loss. In Section \ref{sec4}, we develop an efficient solution to jointly optimize device selection, receiver beamforming, and RIS phase shifts. In Section \ref{sec6}, we present extensive numerical results to evaluate the proposed method. Finally, this paper concludes in Section \ref{sec7}.

\emph{Notation}: Throughout, we use $\Real$ and $\Complex$ to denote the real and complex number sets, respectively. Regular letters, bold small letters, and bold capital letters are used to denote scalars, vectors, and matrices, respectively. We use  $(\cdot)^T$ and $(\cdot)^H$ to denote the transpose and the conjugate transpose, respectively.  We use $x_i$ to denote the $i$-th entry of vector $\xv$, $x_{ij}$ to denote the $(i,j)$-th entry of matrix $\Xv$, $\CN(\mu,\sigma^2)$ to denote the circularly-symmetric complex normal distribution with mean $\mu$ and covariance $\sigma^2$, $\abs{\mathcal{S}}$ to denote the cardinality of set $\mathcal{S}$,
$\norm{\cdot}_p$ to denote the $\ell_p$ norm,  $\Iv_N$ to denote the $N\times N$ identity matrix,  $\diag(\xv)$ to denote a diagonal matrix with the diagonal entries specified by $\xv$, and $\E[\cdot]$ to denote the expectation operator.
\section{System Model}\label{sec2}
\subsection{FL System}\label{sec2a}
\begin{figure}[!t]
	\centering
	\includegraphics[width=4.2in]{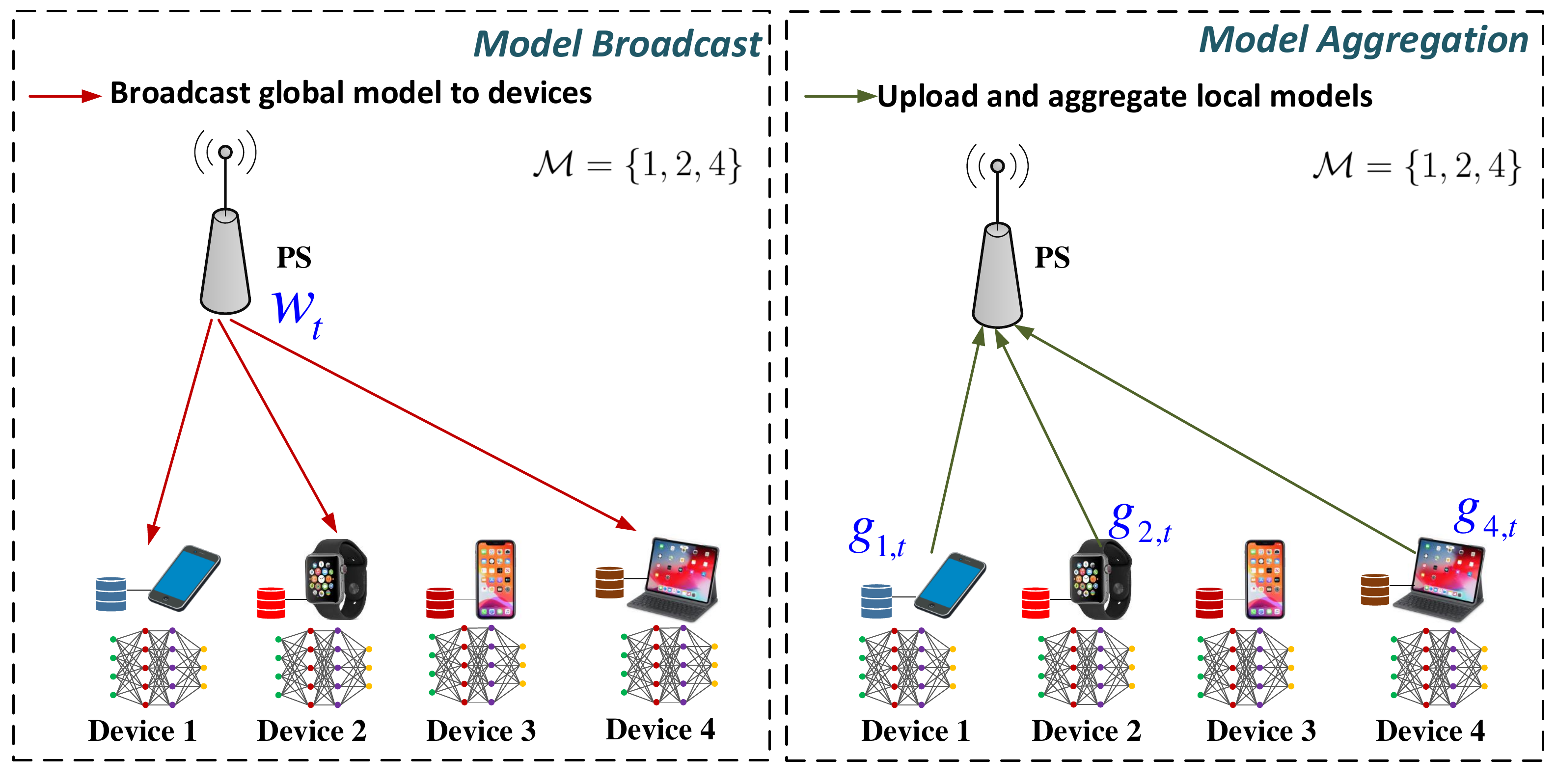}
	\caption{Illustration of the $t$-th FL training round with $M=4$.}
	\label{Fig1}
\end{figure}

We consider a general FL system comprising a PS and $M$ edge devices as depicted in Fig. \ref{Fig1}. The learning objective is to minimize an empirical loss function
\begin{align}\label{eq01}
\min_{\wv\in\Real^{D\times 1}}	F(\wv)=\frac{1}{K} \sum_{k=1}^{K}	f(\wv;\xv_{k},y_{k}),
\end{align}
where $\wv$ is the $D$-dimensional model parameter vector; $K$ is the total number of training samples; $(\xv_k,y_k)$ is the $k$-th training sample with the input feature vector $\xv_k$ and the output label $y_k$; and $f(\wv;\xv_{k},y_{k})$ is the loss function with respect to $(\xv_k,y_k)$. Suppose that the training samples are distributed at the edge devices, and the $m$-th device has $K_m$ training samples with $\sum_{m=1}^MK_m=K$. Denote by $\mathcal{D}_m=\{(\xv_{mk},y_{mk}):1\leq k\leq K_m\}$ the local training dataset at the $m$-th device with $\abs{\mathcal{D}_m}=K_m$. The learning task in  \eqref{eq01} can be represented as
\begin{align}\label{eq02}
	F(\wv)=\frac{1}{K} \sum_{m=1}^{M} K_m F_m(\wv;\mathcal{D}_m) \text{ with } F_m(\wv;\mathcal{D}_m)=\frac{1}{K_m} \sum_{(\xv_{mk},y_{mk})\in\mathcal{D}_m}	f(\wv;\xv_{mk},y_{mk}).
\end{align}

To perform FL, the edge devices update their local models by minimizing $F_m$, and the PS aggregates the global model based on the local updates. In this paper, we adopt batch gradient descent \cite{BGD} for local model update. Specifically, the model $\wv$ is computed iteratively with $T$ training rounds. The $t$-th round, $1\leq t\leq T$, includes the following procedures:
\begin{itemize}
	\item \emph{Device selection}: The PS selects a subset of mobile devices $\mathcal{M}_t\subset \{1,2,\cdots,M\}$ to participate in the learning process. We refer to the devices in $\mathcal{M}_t$ as the active devices.
	
	\item \emph{Model broadcast}: The PS broadcasts the current global model $\wv_t$ to the active devices.
	\item \emph{Local gradient computation}: Each active device computes its local gradient with respect to the local dataset. Specifically, the gradient is given by
	\begin{align}\label{eq03}
\gv_{m,t}\triangleq\nabla F_m(\wv_t;\mathcal{D}_m)\in\Real^D,
	\end{align}
where $\nabla F_m(\wv_t;\mathcal{D}_m)$ is the gradient of $F_m(\cdot)$ at $\wv=\wv_t$.
\item \emph{Model aggregation}: The active devices upload $\{\gv_{m,t}\}$ to the PS {through wireless channels. Based on the received signals, the PS intends to compute a weighted sum of the local gradients $\{\gv_{m,t}:m\in\mathcal{M}_t\}$, which is used to update the global model. In this paper, we consider the popular FL setup introduced in \cite{FEDSGD}, where the weight of the $m$-th local gradient vector is proportional to the size of the corresponding local dataset $K_m$. In this case, we want to estimate   $\rv_t\triangleq\sum_{m\in\mathcal{M}_t} K_m \gv_{m,t}$ at the PS from the received signals. We refer to $\rv_t$ the (true) global gradient vector, and denote the estimate of $\rv_t$ by $\hat \rv_t$.   Due to channel fading and communication noise, there inevitably exist distortions in the estimation process. Therefore, we generally have $\hat \rv_t\neq \rv_t$. After obtaining $\hat \rv_t$, we update $\wv_{t+1}$ by
} 
	\begin{align}\label{eq04}  
\wv_{t+1}=\wv_t-\frac{\lambda}{\sum_{m\in\mathcal{M}_t} K_m}{\hat \rv_t},
\end{align}
where a scaling factor $1/({\sum_{m\in\mathcal{M}_t} K_m})$ is introduced \cite{FEDSGD}; and  $\lambda$ is the learning rate.
\end{itemize}

 To improve the communication efficiency, we select a subset of the devices to be active by following \cite{FEDSGD}. For simplicity, we assume that the wireless channel is invariant during the learning process. The studies on time-varying channels can be found in Section \ref{sec6g}.
 Consequently, we fix the device selection for all iterations, i.e., $\mathcal{M}_t=\mathcal{M}, \forall t$. Note that $\mathcal{M}$ should be jointly optimized with the transceivers by taking into account both the channel heterogeneity (\eg channel fading varies across the edge devices) and the data heterogeneity (\eg the devices hold different numbers of  training samples).

{\remark{There exist many variants of the above gradient descent algorithm. For example, the authors in \cite{FEDSGD} replaced the single gradient $\gv_{m,t}$ in  \eqref{eq03} with a sequence of multiple-batch gradient descent updates (a.k.a. federated stochastic gradient descent) and/or multiple-epoch updates (a.k.a. federated averaging). Although the analysis in this paper adopts the model in  \eqref{eq03}, we show by numerical results in Section \ref{sec6d} that the proposed approach can improve the learning performance of multiple-batch/epoch FL algorithms as well.
}}

\subsection{RIS-Assisted Communication System}\label{sec2b}
\begin{figure}[!t]
	\centering
	\includegraphics[width=4in]{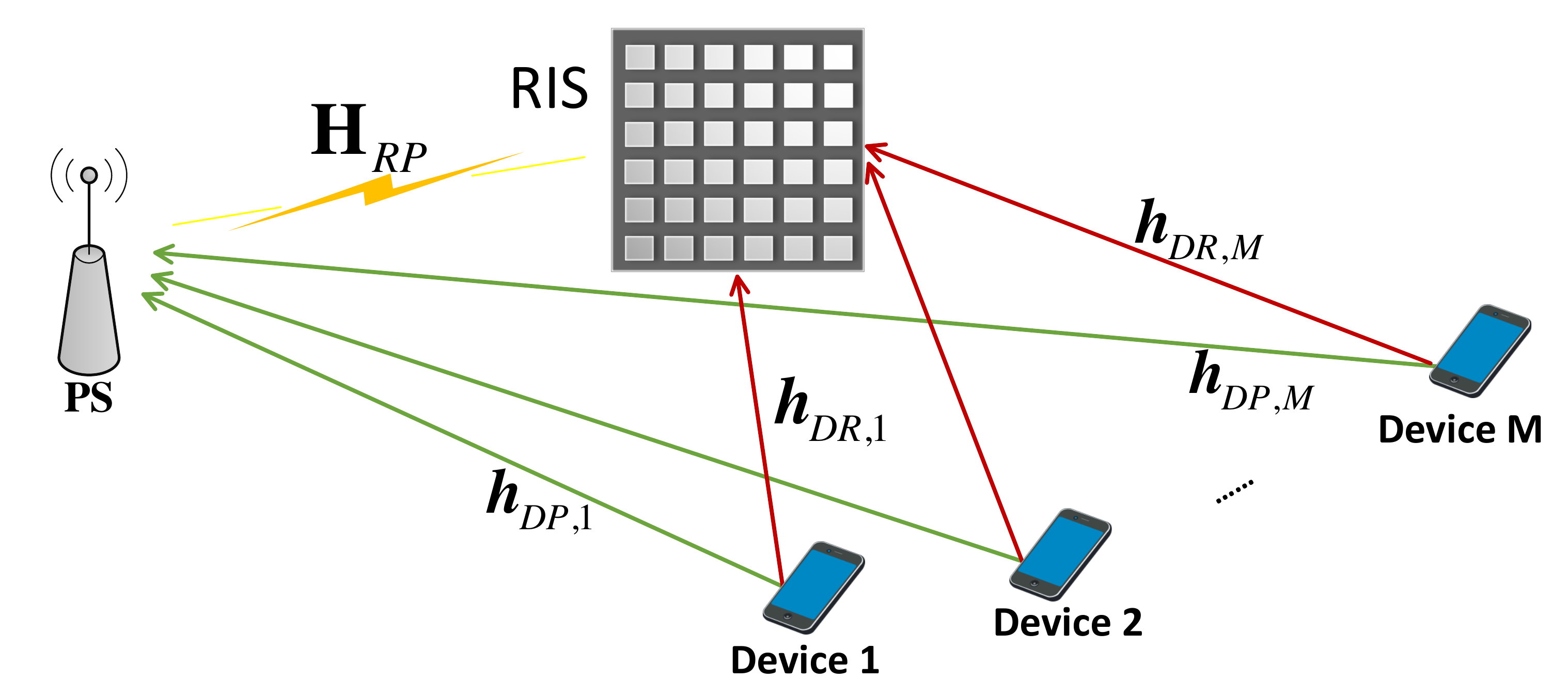}
	\caption{A RIS-assisted communication system.}\label{Fig2}
\end{figure}

The underlying wireless network for the above FL system is depicted in Fig. \ref{Fig2}, where a RIS is deployed to assist the communication between the single-antenna edge devices and the $N$-antenna PS. Suppose that the RIS has $L$ phase shift elements. We assume a block fading channel model, where the channel coefficients remain invariant during the whole FL training process. Let $\hv_{DP,m}\in \Complex^{N\times 1}$, $\Hv_{RP}\in \Complex^{N\times L}$, and $\hv_{DR,m}\in \Complex^{L\times 1}$, $m=1,2,\cdots,M$, denote the direct $m$-th-device-PS, the RIS-PS, and the $m$-th-device-RIS channel coefficient vectors/matrix, respectively. We assume perfect channel state information (CSI) at the PS, the RIS controller, and the devices.\footnote{Studies on CSI acquisition in RIS-assisted systems can be found, \eg in \cite{LIS_He,RIS_Hang,ZWang_CERIS}.}
Thanks to the reconfigurable property of the RIS, the RIS elements induce independent phase shifts on the incident signals.
We keep the RIS phase shifts invariant during the FL model aggregation and denote the RIS phase-shift vector as $\thetav\in \Complex^{L\times 1}$. {We set the RIS amplitude reflection coefficients to $1$ and
assume continuous RIS phase shifts with $\abs{\theta_l}^2=1$ for $l=1,2,\cdots,L$.}
For ease of notation, we define the \emph{effective} $m$-th-device-PS channel coefficient vector as 
\begin{align}\label{eq05}
\hv_m(\thetav)\triangleq \hv_{DP,m}+\Hv_{RP}\diag(\thetav)\hv_{DR,m}=\hv_{DP,m}+\Gv_m\thetav,
\end{align}
where $\Gv_m\triangleq \Hv_{RP}\diag(\hv_{DR,m})\in \Complex^{N\times L}$ is the $m$-th cascaded channel coefficient matrix. 

In the model aggregation of the $t$-th training round, we denote the transmit signal from the active devices in time slot $d$ by $\{x_{m,t}[d]:m\in \mathcal{M}\}$.  The corresponding received signal at the PS, denoted by $\yv_t[d]$, is the superposition of the signals from the direct channels and the device-RIS-PS cascaded channels, \ie
\begin{align}\label{eq06}
	\yv_t[d]&=\sum_{m\in\mathcal{M}} \left( \hv_{DP,m}+\Hv_{RP}\diag(\thetav)\hv_{DR,m}\right) x_{m,t}[d]+\nv_t[d]=\sum_{m\in\mathcal{M}}\hv_m(\thetav) x_{m,t}[d]+\nv_t[d],
\end{align}
where $\nv_t[d]\in \Complex^{N\times 1}$ is an additive white Gaussian noise (AWGN) vector with the entries following the distribution of $\CN(0,\sigma_n^2)$.

\subsection{Over-the-Air Model Aggregation}
We adopt over-the-air computation to achieve FL model aggregation \cite{GZhu_BroadbandAircomp}. Specifically, at the $t$-th training round, $1\leq t\leq T$,  the active devices upload their local model updates $\{\gv_{m,t}:m\in\mathcal{M}\}$ using the same time-frequency resources. By controlling the transmit equalization factors, the desired weighted sum $\hat \rv_t$ in \eqref{eq04} is coherently constructed at the PS. The details are discussed as follows.

In the sequel, we omit the index $t$ for brevity.  
Denote the $d$-th entry of $\gv_{m}$ by $g_m[d]$.
To perform over-the-air model aggregation, $\{g_{m}[d]\}$ are first transferred to $D$-slot transmit signals $\{x_{m}[d]: 1\leq d\leq D,m\in\mathcal{M}\}$. Specifically, we first compute the local gradient statistics (\ie the means and variances) for $\forall m\in\mathcal{M}$ by
\begin{align}
	&\bar g_{m}=\frac{1}{D}\sum_{d=1}^D g_{m}[d],
	\quad \nu_{m}^2=\frac{1}{D}\sum_{d=1}^D (g_{m}[d]-\bar g_{m})^2.\label{eq08}
\end{align}
Then, the active devices upload these scalar quantities $\{\bar g_{m},\nu_m^2\}$ to the PS.\footnote{ The scalars $\{\bar g_{m},\nu_m^2\}$ can be transmitted to the PS via conventional OMA techniques, which requires an overhead proportional to $|\mathcal{M}|$. Since a typical learning model contains thousands or even millions of parameters (\ie $D\gg |\mathcal{M}|$), we assume that the uploading of  $\{\bar g_{m},\nu_m^2\}$ is error-free with a neglectable overhead.}

The $m$-th device sets the transmit sequence  $\{x_{m}[d]: 1\leq d\leq D\}$ as
\begin{align}
x_{m}[d]=p_{m}s_m[d] \text{ with } s_m[d]\triangleq \frac{g_{m}[d]-\bar g_{m}}{\nu_{m}}, \forall d,\label{eq10}
\end{align}
where $p_{m}\in \Complex$ is the transmit equalization factor used to combat the channel fading and achieve the desired weight (\ie $K_m$) at the PS. {In \eqref{eq10}, $g_m[d]$ is  mapped to a zero-mean unit-variance symbol $s_m[d]$ by using $\bar g_m$ and $\nu_m$ such that $\E[s_m[d]]=0$ and $\E[s_m^2[d]]=1$.} This normalization step ensures that $\E[\abs{x_{m}[d]}^2]=\abs{p_{m} }^2$, where $\abs{p_{m} }^2$ controls the transmit power. Here, we consider an individual transmit power constraint as
\begin{align}
	\E\left[\abs{x_{m}[d]}^2\right]=\abs{p_{m} }^2\leq P_0,\forall m,\label{eq11}
\end{align} 
where $P_0$ is the maximum transmit power. 

Substituting \eqref{eq10} into \eqref{eq06}, the received signal at the PS in time slot $d$ is given by
\begin{align}\label{eq12}
	\yv[d]&=\sum_{m\in\mathcal{M}}\hv_m(\thetav)\frac{ p_{m}}{\nu_{m}}(g_{m}[d]-\bar g_{m})+\nv[d].
\end{align}
The PS computes the estimate of $r[d]=\sum_{m\in\mathcal{M}} K_m g_{m}[d]$, \ie $\hat r[d]$, from   $\yv[d]$ by a linear estimator as
\begin{align}\label{eq13}
	\hat r[d]=\frac{1}{\sqrt{\eta}}\fv^H\yv[d]+\bar g=\frac{1}{\sqrt{\eta}}\left( \sum_{m\in\mathcal{M}}\fv^H\hv_m(\thetav)\frac{ p_{m}}{\nu_{m}}(g_{m}[d]-\bar g_{m})+\fv^H\nv[d]\right) +\bar g,
\end{align}
where $\bar g\triangleq\sum_{m\in\mathcal{M}} K_m \bar g_{m}$;
$\fv\in\Complex^{N\times 1}$ is the normalized receiver beamforming vector with $\norm{\fv}_2^2=1$; and $\eta>0$ is a normalization scalar. {In \eqref{eq13}, we add an additional term $\bar g=\sum_m K_m\bar g_m$ that is related to the local means $\{\bar g_m\}$ since $\bar g_m$ has been subtracted in the normalization step in \eqref{eq10}. }

After estimating $\hat \rv=[\hat r[1],\cdots,\hat r[D]]$, we update the global model by \eqref{eq04}. Note that {the existences of fading and communication noise} lead to inevitable estimation error in $\hat \rv$. As a result, the global model update in \eqref{eq04} becomes inaccurate, which affects the convergence of FL.
 In the next section, we quantitatively characterize the impact of the communication error on the convergence of FL with respect to $\{\mathcal{M},\fv,\thetav,\eta,p_m\}$.
{\remark{We note that the proposed over-the-air model aggregation framework is different from the existing solutions  \cite{FL_2,FL_1,GZhu_BroadbandAircomp} in the following two aspects:
		\begin{itemize}
			\item Refs. \cite{FL_2,FL_1} only assume that $g_m[d]$ is mapped to $s_m[d]$ without discussing the details on the normalization method. Besides, Ref. \cite{GZhu_BroadbandAircomp} sets $s_m[d]={(g_{m}[d]-\bar g)}/{\nu}$, where $\bar g$ is defined in \eqref{eq13} and $\nu^2\triangleq \sum_{m\in\mathcal{M}} K_m \nu^2_m$. In other words, two uniform statistics $\bar g$ and $\nu^2$ are used to compute $s_m[d]$ for $\forall m$. When the local gradients $\{\gv_m\}$ significantly vary among the active devices, this normalization cannot guarantee $\E[s_m^2[d]]=1$ for $\forall m$, and hence the constraint in  \eqref{eq11} may become unsatisfied. In contrast, we normalize the transmit signal by the local gradient statistics, ensuring that \eqref{eq11} always holds.
			\item Refs. \cite{FL_2,FL_1,GZhu_BroadbandAircomp} measure the over-the-air model aggregation performance by the estimation error with respect to $\sum_{m\in\mathcal{M}} K_m s_{m}[d]$ rather than that of $\sum_{m\in\mathcal{M}} K_m g_{m}[d]$. In contrast, we directly consider the estimation on $\hat r[d]$ in \eqref{eq13} by explicitly taking the de-normalization step into account. By doing so, we can model the impact of the whole model aggregation process on the learning performance as detailed in Section \ref{sec3}. 
		\end{itemize}
}}
%
%

\section{FL Performance Analysis and Problem Formulation}\label{sec3}
In this section, we analyze the performance of the FL task under the over-the-air model aggregation framework. In Section \ref{sec3a}, we introduce several assumptions on the loss function $F(\cdot)$. Based on these assumptions, we derive a tractable upper bound on the learning loss $F(\wv_t)$ by analyzing the model updating error caused by the device selection and the communication noise. Then, we give a closed-form solution to $\{\eta,p_m,\forall m\}$ that minimizes the loss bound.
Finally, we formulate the system design problem as a unified learning loss minimization task over the device selection decision $\mathcal{M}$, the receiver beamforming vector $\fv$, and the RIS phase shift vector $\thetav$.

\subsection{Assumptions and Preliminaries}\label{sec3a}
Recall that the PS computes $\hat \rv_t$ by the over-the-air model aggregation in \eqref{eq13}. Consequently, the global model update recursion is given by
\begin{align}\label{eq17}
	\wv_{t+1}=\wv_t-\frac{\lambda}{\sum_{m\in\mathcal{M}} K_m}\hat \rv_t=\wv_t-\lambda (\nabla F(\wv_t)-\ev_t ),
\end{align}
where $\nabla F(\wv_t)\triangleq \frac{1}{K}\sum_{k=1}^K\nabla f(\wv_t;\xv_{k},y_{k})$ is the gradient of $F(\wv)$ at $\wv=\wv_t$; and $\ev_t$ denotes the gradient error vector due to the device selection and model aggregation. Specifically, $\ev_t$ is given by
\begin{align}\label{eq18}
	\ev_t=&\underbrace{\nabla F(\wv_t)-\frac{ \rv_{t}}{\sum_{m\in\mathcal{M}}  K_m}}_{ \ev_{1,t}\text{(error due to device selection)}}+\underbrace{\frac{ \rv_{t}}{\sum_{m\in\mathcal{M}}  K_m}-\frac{ \hat \rv_t}{\sum_{m\in\mathcal{M}}  K_m}}_{ \ev_{2,t}\text{(error due to fading and communication noise)}}.
\end{align}
{From \eqref{eq18}, we see that $\ev_{1,t}=\bf 0$ if $\mathcal{M}=\{1,2,\cdots,M\}$ and $\ev_{1,t}\neq \bf 0$ if $|\mathcal{M}|<M$. Intuitively, device selection reduces the amount of exploited data and induces additional error $\ev_{1,t}$ on the gradient vector $\nabla F(\wv_t)$.  As shown later in this section, this error, together with the error in model aggregation (\ie $\ev_{2,t}$), jointly jeopardizes the performance of FL.}

To proceed, we make the following assumptions on the loss function $F(\cdot)$:
%
\begin{description}
\item[A1] $F$ is strongly convex with parameter $\mu$. That is,
	$F(\wv)\geq F(\wv^\prime)+(\wv-\wv^\prime)^T\nabla F(\wv^\prime)+\frac{\mu}{2}\norm{\wv-\wv^\prime}_2^2, \forall \wv,\wv^\prime\in\Real^{D\times 1}$.
\item[A2] The gradient $\nabla F(\cdot)$ is Lipschitz continuous with parameter $\omega$. That is, 
$\norm{\nabla F(\wv)-\nabla F(\wv^\prime)}_2\leq \omega\norm{\wv-\wv^\prime}_2,\forall \wv,\wv^\prime\in\Real^{D\times 1}$.
\item[A3] $F$ is twice-continuously differentiable.
\item[A4] The gradient with respect to any training sample, $\nabla f(\cdot)$, is upper bounded at $\{\wv_t:\forall 1\leq t\leq T\}$. That is,
\begin{align}
	\norm{\nabla f(\wv_{t};\xv_{k},y_{k})}_2^2\leq \a_1+\a_2 \norm{\nabla F(\wv_{t})}_2^2, \forall 1\leq k\leq K,\forall 1\leq t\leq T,\label{eq20}
\end{align}
for some constants $\a_1\geq 0$ and $\a_2> 0$.
\end{description}
{Assumptions A1--A4 are standard  in the stochastic optimization literature; see, \eg \cite{NDP,Friedlander2012,chen2019joint}.} 
Assumption A1 ensures that a global optimum $\wv^\star$ exists for the loss function $F$. {Assumption A4 provides a bound on the norm of the local gradient vectors. Note that the parameter $\alpha_1$ allows the local gradient to be nonzero even if the global gradient is zero. In practice, Assumptions A1-A3 hold for many regularized problems; see, \eg \cite[Section 5]{Friedlander2012}. Concerning A4, it is reasonable to expect that $K,T<\infty$ and $\norm{\nabla f(\wv_{t};\cdot)}_2^2<\infty$ in practice. In this case, we can exhaustively check the values of $\{\norm{\nabla f(\wv_{t};\xv_{k},y_{k})}_2^2\}_{k=1}^K$ over $\{\wv_t\}_{t=1}^T$ to find sufficiently large $\a_1$ and $\a_2$ such that A4 holds.}

As shown in \cite[Lemma 2.1]{Friedlander2012}, Assumptions A1--A4 lead to an upper bound on the loss function $F(\wv_{t+1})$ with respect to the recursion \eqref{eq17} with a proper choice of the learning rate $\lambda$. The details are given in the following lemma.
\lemma{\label{lemma1} Suppose that $F(\cdot)$ satisfies Assumptions A1--A4. At the $t$-th training round, with $\lambda=1/\omega$,  we have
	\begin{align}
		\E[F(\wv_{t+1})]\leq \E[F(\wv_{t})]-\frac{1}{2\omega }\norm{\nabla F(\wv_t)}_2^2+\frac{1}{2\omega}\E[\norm{\ev_t}_2^2],
	\end{align}
where {the Lipschitz constant $\omega$ is defined in Assumption A2}, and the expectations are taken with respect to the communication noise. 
} 
\proof{See \cite[Lemma 2.1]{Friedlander2012}.}
\subsection{Learning Performance Analysis}\label{sec3b}
With Assumptions A1--A4 and Lemma \ref{lemma1}, we are ready to present the performance analysis. Specifically, we apply Lemma \ref{lemma1} and bound the term $\E[\norm{\ev_t}^2]$ to derive a tractable expression of $	\E[F(\wv_{t+1})]$. This expression is further used to obtain an upper bound of the average difference between the training loss and the optimal loss, \ie $\E\left[F(\wv_{t+1})-F(\wv^\star)\right]$.

To begin with, we apply Lemma \ref{lemma1} and obtain
\begin{align}\label{temp04}
	\E[F(\wv_{t+1})]&\leq \E[F(\wv_{t})]-\frac{1}{2\omega}\norm{\nabla F(\wv_t)}_2^2+\frac{1}{2\omega}\E[\norm{\ev_{1,t}+\ev_{2,t}}_2^2]\nonumber\\
	&\overset{(a)}\leq \E[F(\wv_{t})]-\frac{1}{2\omega}\norm{\nabla F(\wv_t)}_2^2+\frac{1}{2\omega}\E\left[ (\norm{\ev_{1,t}}_2+\norm{\ev_{2,t}}_2)^2\right]\nonumber\\
	 &\overset{(b)}\leq E[F(\wv_{t})]-\frac{1}{2\omega}\norm{\nabla F(\wv_t)}_2^2+\frac{1}{\omega} (\norm{\ev_{1,t}}^2_2+\E\left[\norm{\ev_{2,t}}^2_2\right])
\end{align}
where {the expectations are taken with respect to the communication noise; $(a)$ is from the triangle inequality; and $(b)$ is from the inequality of arithmetic and geometric means (AM–GM inequality). In the last step of \eqref{temp04}, we drop the expectation on $\norm{\ev_{1,t}}^2_2$ since $\ev_{1,t}$ is independent to the communication noise.}

To analyze $\E[F(\wv_{t+1})]$, we need tractable expressions for $\norm{\ev_{1,t}}_2^2/\omega$ and $\E[\norm{\ev_{2,t}}_2^2]/\omega$. Note that the error $\ev_{1,t}$ is determined by the device selection decision $\mathcal{M}$, whereas the communication error $\ev_{2,t}$ is related to $\mathcal{M}$ and  $\{\fv,\thetav,\eta,p_m,\forall m\in\mathcal{M}\}$. For given $\mathcal{M}$, the choice of $\{\eta,p_m,\forall m\}$ that minimizes $\E[\norm{\ev_{2,t}}_2^2]$ is given in the following lemma.

{\lemma{\label{lemmaa}Under the transmit power constraint \eqref{eq11}, the optimal $\{\eta,p_{m}\}$ that minimizes $\E[\norm{\ev_{2,t}}_2^2]$ is given by
			\begin{align}
				& \eta=\min_{m\in\mathcal{M}} \frac{P_0\abs{\fv^H\hv_m(\thetav)}^2}{K_m^2 \nu_m^2}, p_{m}=\frac{K_m \sqrt{\eta} \nu_m(\fv^H\hv_m(\thetav))^H}{\abs{\fv^H\hv_m(\thetav)}^2}.\label{eq15}
			\end{align}
		Moreover, $\E[\norm{\ev_{2,t}}_2^2]$ (as a function of $\mathcal{M}$, $\thetav$, and $\fv$) with \eqref{eq15} is given by
		\begin{align}\label{eq16}
			\E[\norm{\ev_{2,t}}_2^2]=\frac{D\sigma^2_n}{P_0\left(	\sum_{m\in \mathcal{M}}K_m\right)^2}  \max_{m\in\mathcal{M}}  \frac{K_m^2\nu_{m}^2 }{\abs{\fv^H\hv_m(\thetav)}^2}.
		\end{align}
}}
\proof{See Appendix \ref{appa0}.}

Lemma \ref{lemmaa} gives the optimal solution to $\{\eta,p_m,\forall m\}$ and the corresponding expression of $\E[\norm{\ev_{2,t}}_2^2]$. Based on Lemma \ref{lemmaa}, we bound $\E\left[F(\wv_{t+1})-F(\wv^\star)\right]$ for given $\{\mathcal{M}$, $\fv$, $\thetav\}$ in the following theorem.

\theorem{\label{the1} With Assumptions A1--A4, $\lambda=1/\omega$, and $\{\eta,p_m\}$ given in \eqref{eq15}, for any $\{\mathcal{M}$, $\fv$, $\thetav\}$ and $t=0,1,\cdots,T-1$, we have
	\begin{align}\label{eq23}
		\E\left[F(\wv_{t+1})-F(\wv^\star)\right]	\leq \frac{\a_1}{\omega}d(\mathcal{M},\fv,\thetav)\frac{1-\left(\Psi(\mathcal{M},\fv,\thetav)\right)^t }{1-\Psi(\mathcal{M},\fv,\thetav) }+\left(\Psi(\mathcal{M},\fv,\thetav)\right)^t (F(\wv_0)-F(\wv^\star)),
	\end{align}
	where $\wv_0$ is the initial FL model; {and the functions  $d(\mathcal{M},\fv,\thetav)$ and $\Psi(\mathcal{M},\fv,\thetav)$ are defined as}
	\begin{align}
		&d(\mathcal{M},\fv,\thetav){\triangleq}\frac{4}{K^2} \left(K-\sum_{m\in \mathcal{M}}K_m\right)^2+\frac{\sigma_n^2}{P_0\left(\sum_{m\in \mathcal{M}}K_m\right)^2} \max_{m\in \mathcal{M}} \frac{K_m^2}{\abs{\fv^H\hv_m(\thetav)}^2	},		\label{eq24}	\\
		&\Psi(\mathcal{M},\fv,\thetav){\triangleq}1-\frac{\mu}{\omega}+\frac{2\mu\a_2d(\mathcal{M},\fv,\thetav)}{\omega}.\label{eq25}
	\end{align}
{In the above, the parameters $\{\mu,\omega,\a_1,\a_2\}$ are defined in Assumptions A1--A4; $P_0$ is defined in \eqref{eq11}; and $\sigma_n^2$ is defined in \eqref{eq06}.
}
}
\proof{See Appendix \ref{appa}.}

From Theorem \ref{the1}, we see that $\E\left[F(\wv_{t+1})-F(\wv^\star)\right]$ is upper bounded by a quantity related to $\mathcal{M}$, $\fv$, and $\thetav$. Moreover, this upper bound converges with speed $\Psi(\mathcal{M},\fv,\thetav)$ when $\Psi(\mathcal{M},\fv,\thetav)<1$. The following corollary further characterizes  the convergence behavior of  $\E\left[F(\wv_{t+1})-F(\wv^\star)\right]$.

\corollary{\label{cor1}Suppose that the assumptions in Theorem \ref{the1} hold. Also, suppose that $d(\mathcal{M},\fv,\thetav)\leq \frac{1}{2\a_2}$. As $T\to \infty$, we have
	\begin{align}\label{eq26}
		\lim_{T\to \infty}\E\left[F(\wv_{T})-F(\wv^\star)\right]	\leq \frac{\a_1d(\mathcal{M},\fv,\thetav)}{\omega-\mu+2\mu\a_2d(\mathcal{M},\fv,\thetav)}.
	\end{align}
}
\proof{When $d(\mathcal{M},\fv,\thetav)\leq \frac{1}{2\a_2}$, we have $\Psi(\mathcal{M},\fv,\thetav)<1$. Therefore, $\lim_{T\to\infty} (\Psi(\mathcal{M},\fv,\thetav))^T=0$. Plugging this result into \eqref{eq23}, we obtain \eqref{eq26}.
}
\subsection{Problem Formulation}

Corollary \ref{cor1} shows that the FL recursion \eqref{eq17} guarantees to converge with a sufficiently small $d(\mathcal{M},\fv,\thetav)$. However, there generally exists a gap between the converged loss $\lim_{T\to \infty}\E\left[F(\wv_{T})\right]$ and the optimal one $E\left[F(\wv^\star)\right]$ because of the device selection loss and the communication noise. Moreover, the upper bound of this performance gap is a monotonic function of  $d(\mathcal{M},\fv,\thetav)$ as shown in the right hand side of \eqref{eq26}. Besides, as shown in \eqref{eq25}, the convergence rate $\Psi(\mathcal{M},\fv,\thetav)$ is also a  monotonic function of  $d(\cdot)$. From Theorem \ref{the1} and Corollary \ref{cor1}, we conclude that the quantity $d(\cdot)$ in \eqref{eq24} represents the impact of the device selection and the communication error on the convergence rate and the asymptotic learning performance. Specifically, a smaller  $d(\cdot)$ leads to faster convergence and a smaller gap in $\lim_{T\to \infty}\E\left[F(\wv_{T})-F(\wv^\star)\right]$. This motivates us to treat $d(\cdot)$ as the metric of the FL performance and optimize $\{\mathcal{M},\fv,\thetav\}$ by minimizing $d(\mathcal{M},\fv,\thetav)$.\footnote{Note that $d(\mathcal{M},\fv,\thetav)$ is independent to the values of $\{\omega,\mu,\a_1,\a_2\}$. Therefore, our problem formulation is applicable to the general FL loss function that satisfies Assumptions A1--A4, and we do not need to estimate the values of $\{\omega,\mu,\a_1,\a_2\}$ in our design.
} 

To facilitate the problem formulation, we describe the device selection decision $\mathcal{M}$ by a binary indicator vector $\xv=[x_1,\cdots,x_M]\in \{0,1\}^M$ as $x_m=1$ for $m\in \mathcal{M}$ and $x_m=0$ otherwise.
We then formulate the communication-learning design problem as minimizing $d(\mathcal{M},\fv,\thetav)$ over the feasible set of $\{\xv,\fv,\thetav\}$ as\footnote{In the rest of this paper, we use $\mathcal{M}$ and $\xv$ interchangeably.}
\begin{subequations}\label{eq28}
\begin{align}
\min_{\xv,\fv,\thetav}\quad &\frac{4}{K^2} \left(\sum_{m=1}^M (1-x_m)K_m\right)^2+\frac{\sigma_n^2 }{P_0\left(\sum_{m=1}^M {x_m}K_m\right)^2} \max_{m:x_m=1}\frac{K_m^2}{\abs{\fv^H\hv_m(\thetav)}^2}	\label{eq28a}\\
	\text{s.t. }& x_m\in \{0,1\},1\leq m\leq M,\label{eq28b} \norm{\fv}_2^2=1, \abs{\theta_l}^2=1,1\leq l\leq L.
\end{align}
\end{subequations}
{\remark{\label{remark3}Theorem \ref{the1} and Corollary \ref{cor1} are derived under Assumptions A1--A4. Note that the practical learning task may have a loss function that violates these assumptions. For example, the cross-entropy function that is widely-adopted in deep learning is not strongly-convex \cite{DL_Goodfellow}. The results in Theorem \ref{the1} and Corollary \ref{cor1} can not be applied to this case. However, we show by numerical simulations in Section \ref{sec6} that $d(\mathcal{M},\fv,\thetav)$ is still a good metric and the solution to \eqref{eq28} yields a considerable  performance improvement compared with the state-of-the-art approach even without imposing some of the assumptions.
}}
\subsection{Further Discussions}\label{sec3c}
In this subsection, we take a close look at problem \eqref{eq28} and discuss the  important communication-learning tradeoff characterized by the proposed formulation. Besides, we compare our formulation with those in the existing literature and highlight the necessity of introducing the RIS in the considered FL system.

To begin with, we note the tradeoff between learning and communication concerning the number of selected devices, as implied by \eqref{eq28a}. On one hand, selecting more active devices (\ie a larger $\mathcal{M}$) exploits more training data and thus enhances the learning performance \cite{FEDSGD}. Indeed, a larger $\mathcal{M}$ leads to a smaller $4 (\sum_{m=1}^M (1-x_m)K_m)^2/K^2$ and a larger $(\sum_{m=1}^M {x_m} K_m)^2$, both deceasing the objective in \eqref{eq28a}. On the other hand, the communication error, as shown in the second term of \eqref{eq28a}, \ie $\max_{m:x_m=1}{ K_m^2}/{\abs{\fv^H\hv_m(\thetav)}^2}$, is dominated by the active device {with the lowest value of ${\abs{\fv^H\hv_m(\thetav)}^2}/{ K_m^2}$, which depends on the channel conditions}. 
Consequently, selecting more active devices enlarges the communication error, which limits the learning performance. 

The existing work on wireless FL \cite{GZhu_BroadbandAircomp} has made an initial attempt to analyze this communication-learning tradeoff, where the device selection loss is measured by the fraction of the exploited data. Furthermore, motivated by this tradeoff, the authors in \cite{FL_1} proposed to maximize the number of active devices (\ie $\abs{\mathcal{M}}$) while constraining the communication mean-square error (MSE) below some predetermined threshold.  In other words, the loss subject to the device selection is approximately measured by $\abs{\mathcal{M}}$, and it is hoped that a proper choice of the MSE threshold well represents the communication-learning tradeoff. Compared with the frameworks in \cite{GZhu_BroadbandAircomp,FL_1}, our proposed formulation has the following two advantages:
\begin{itemize}
	\item The objective in \eqref{eq28a} directly unifies the device selection loss and the communication error in assessing the FL performance, and thus quantitatively characterizes the communication-learning tradeoff.  In contrast,  Refs. \cite{GZhu_BroadbandAircomp,FL_1} use either the fraction of exploited data or the number of active devices as a rough proxy of the learning performance. Besides, we do not need to tune the MSE threshold like in \cite{FL_1}.
	\item The proposed formulation characterizes the effect of varying dataset sizes (\ie $\{K_m\}$). Specifically, discarding a device with a large $K_m$ is unfavorable since this significantly increases the first term in  \eqref{eq28a}. Therefore, the optimal device selection decision should balance exploiting more data and reducing the communication error.  
	In contrast, the metric used in \cite{FL_1} (\ie the number of active devices  $\abs{\mathcal{M}}$) equally treats all the devices and thus is unable to adapt to the characteristics of a particular dataset.
\end{itemize}

As a final remark, we note that the RIS phase shifts $\thetav$ is critical in learning performance optimization. As discussed above, in order to fully optimize the learning performance, it is necessary to select as many devices as possible and at the same time to reduce the communication MSE as much as possible. However, due to random channel conditions, selecting a large number of devices inevitably degrades the communication performance, which cannot be mitigated by only relying on the receiver beamforming $\fv$ in a large system. In this case, tuning the RIS phase shifts can enhance the weak channels and break this dilemma.

\section{Communication-Learning Co-Design}\label{sec4}
As discussed in the previous section, the communication variables (\ie the receiver beamforming and the RIS phase shifts) and the learning design (\ie the device selection) should be jointly optimized to maximize the learning performance in \eqref{eq28}. 
However, Problem \eqref{eq28} is a mixed-integer non-convex optimization problem that involves combinatorial search over $\xv$ and the non-convex objective/constraints.
In this section, we propose an iterative Gibbs-sampling-based algorithm to optimize the device selection $\xv$, the receiver beamforming $\fv$, and the RIS phase shifts $\thetav$ based on \eqref{eq28}. Our design avoids the high computational complexity of the combinatorial search. We first propose our design on optimizing $\fv$ and $\thetav $ with a given $\xv$ in Section \ref{sec4a}. Then, the device selection decision vector $\xv$ is optimized by Gibbs sampling \cite{PowerControl_Qian} in Section \ref{sec4b}. Finally, we summarize the overall algorithm and discuss its computational complexity.

\subsection{Optimizing Receiver Beamforming and RIS Phase Shifts Given Device Selection}\label{sec4a}
Suppose that the binary variable $\xv$ (or equivalently, $\mathcal{M}$) is given in \eqref{eq28}. We can simplify \eqref{eq28} as
	\begin{align}
		{\min_{(\fv,\thetav)\in\mathcal{S}}} & {\max_{m\in \mathcal{M}}} u_m(\fv,\thetav)\triangleq-\frac{\abs{\fv^H\hv_m(\thetav)}^2}{ K_m^2},	\label{eq29}
	\end{align}
where $\mathcal{S}=\{\fv\in\Complex^{N\times 1},\thetav\in\Complex^{L\times 1}:\norm{\fv}_2^2=1,\abs{\theta_l}^2=1,1\leq l\leq L\}$ is the feasible  set of $\fv$ and $\thetav$.
Even with a given $\xv$, Problem \eqref{eq29} is still non-convex because of the coupling of $\fv$ and $\thetav$. A natural way to address the problem is to alternately optimize $\fv$ and $\thetav$. For example, Ref. \cite{FL_RIS2} adopts this alternating optimization principle to solve a similar problem, where $\fv$ and $\thetav$ are sequentially optimized by matrix lifting and difference-of-convex (DC) programming until convergence. However, alternating optimization may suffer from slow convergence. Moreover, the matrix lifting technique exhibits high computational cost because  1) it squares the number of variables; and 2) it requires additional decomposition procedures (\eg Cholesky decomposition) to round the solution back.

Instead of the alternating optimization technique, we employ the SCA principle \cite{SCA} to iteratively update $\fv$ and $\thetav$ by solving a sequence of convex  approximate problems.
At iteration $i$, $i=1,2,\cdots$, a convex surrogate function $\widetilde{u}_m^{(i)}(\fv,\thetav)$ is constructed based on $(\fv^{(i)},\thetav^{(i)})\in\mathcal{S}$. The surrogate function can be viewed as an approximation of $u_m(\fv,\thetav)$, which is used to obtain $\fv^{(i+1)}$ and $\thetav^{(i+1)}$. Following \cite[Section V]{SCA}, we set 
\begin{align}
	\widetilde u_m^{(i)}&(\fv,\thetav)=u_m(\fv^{(i)},\thetav^{(i)})+\Re\left\{(\fv-\fv^{(i)})^H\nabla_{\fv^\star}u(\fv^{(i)},\thetav^{(i)})\right\}+\frac{\tau}{K_m^2}\norm{\fv-\fv^{(i)}}^2\nonumber\\
	&+\Re\left\{(\thetav-\thetav^{(i)})^H\nabla_{\thetav^\star}u(\fv^{(i)},\thetav^{(i)})\right\}+\frac{\tau}{K_m^2}\norm{\thetav-\thetav^{(i)}}^2\nonumber\\
	=&\frac{1}{K_m^2}\left( \tau\norm{\fv-\fv^{(i)}}^2+\tau\norm{\thetav-\thetav^{(i)}}^2-2\Re\left\{\fv^H\hv_m(\thetav^{(i)})\hv_m(\thetav^{(i)})^H\fv^{(i)}+\thetav^H\Gv_m^H\fv^{(i)}(\fv^{(i)})^H\hv_m(\thetav^{(i)})\right\}\right) \nonumber\\
	&+\frac{{\abs{(\fv^{(i)})^H\hv_m(\thetav^{(i)})}^2}+2\Re\{(\fv^{(i)})^H\hv_m(\thetav^{(i)})(\thetav^{(i)})^H\Gv_m^H\fv^{(i)}\}}{K_m^2},\label{eq30}
\end{align}
where $\Re\{x\}$ is the real part of $x$; $\nabla_{\fv^\star}u(\fv^{(i)},\thetav^{(i)})$ (or $\nabla_{\thetav^\star}u(\fv^{(i)},\thetav^{(i)})$) is the conjugate gradient of $u(\fv,\thetav)$ with respect to $\fv$ (or $\thetav$) at $\fv=\fv^{(i)}$ and $\thetav=\thetav^{(i)}$; and the regularization parameter $\tau$ is used to ensure strong convexity.

With \eqref{eq30}, $\fv^{(i+1)}$ and $\thetav^{(i+1)}$ are obtained by
	$(\fv^{(i+1)},\thetav^{(i+1)})=	{\argmin_{(\fv,\thetav)\in\mathcal{S}} \max_{m\in \mathcal{M}}} \widetilde u_m^{(i)}(\fv,\thetav)$.
Since $\norm{\fv}_2^2=1$ and $\norm{\thetav}_2^2=L$ for $\forall (\fv,\thetav)\in\mathcal{S}$, the above problem is equivalent to 
\begin{subequations}
	\label{eq32}
	\begin{align}
	(\fv^{(i+1)},\thetav^{(i+1)})= \mathop{\arg\min}_{(\fv,\thetav)\in\mathcal{S},\kappa\in\Real}&\quad  \kappa	\\
		\text{s.t. }& K_m^2\kappa\geq c_m^{(i)}-2\Re\{\fv^H\av_m^{(i)}+\thetav^H\bv_m^{(i)}\},m\in \mathcal{M},\label{eq32b}
	\end{align}
\end{subequations}
where  $\av_m^{(i)}\triangleq\left( \tau+\hv_m(\thetav^{(i)})\hv_m(\thetav^{(i)})^H\right)\fv^{(i)} $; $\bv_m^{(i)}\triangleq\tau\thetav^{(i)}+\Gv_m^H\fv^{(i)}(\fv^{(i)})^H\hv_m(\thetav^{(i)}) $; and $c_m^{(i)}\triangleq\abs{(\fv^{(i)})^H\hv_m(\thetav^{(i)})}^2+2\tau(L+1)+2\Re\{(\fv^{(i)})^H\hv_m(\thetav^{(i)})(\thetav^{(i)})^H\Gv_m^H\fv^{(i)}\}$.

{The problem in \eqref{eq32} is still non-convex since $\mathcal{S}$ is not convex. To tackle this challenge, we \emph{approximately} solve it by solving its Lagrange dual problem. Specifically, define $\zetav=[\zeta_1,\cdots,\zeta_{\abs{\mathcal{M}}}]\geq \bf 0$ as the Lagrange multiplier vector. We define the Lagrangian function of \eqref{eq32} as
	\begin{align}
		\mathcal{L}^{(i)}&(\fv,\thetav,\kappa,\zetav)=\kappa+\sum_{m\in\mathcal{M}}\zeta_m\left( c_m^{(i)}-K_m^2\kappa-2\Re\{\fv^H\av_m^{(i)}+\thetav^H\bv_m^{(i)}\}\right) \nonumber\\
		&=\sum_{m\in\mathcal{M}}\zeta_m c_m^{(i)}+\left(1-\sum_{m\in\mathcal{M}}K_m^2\zeta_m \right) \kappa-2\Re\left\{\fv^H\left( \sum_{m\in\mathcal{M}}\zeta_m\av_m^{(i)}\right) +\thetav^H\left( \sum_{m\in\mathcal{M}}\zeta_m\bv_m^{(i)}\right) \right\}.\label{eqappc01}
	\end{align}
	Then, we solve the dual problem $\max_{\zetav\geq \bf 0}\min_{(\fv,\thetav)\in\mathcal{ S},\kappa\in \Real}\mathcal{L}^{(i)}(\fv,\thetav,\kappa,\zetav)$ to obtain $(\fv^{(i+1)},\thetav^{(i+1)})$. We show in Appendix \ref{appb} that the solution is given by
			\begin{align}
		&\fv^{(i+1)}(\zetav)=\frac{\sum_{m\in\mathcal{M}}\zeta_m\av_m^{(i)}}{\norm{\sum_{m\in\mathcal{M}}\zeta_m\av_m^{(i)}}_2}\text{ and }\thetav^{(i+1)}(\zetav)=e^{\jmath\arg\{\sum_{m\in\mathcal{M}}\zeta_m\bv_m^{(i)}\}},\label{eq33}
	\end{align}
where $\jmath\triangleq \sqrt{-1}$; $\arg \{x\}$ is the argument of complex number $x$ and is applied element-wisely in \eqref{eq33}; and $\zetav$ is given by solving the following convex problem:
\begin{subequations}
	\label{eq34}
	\begin{align}
		\argmin_{\zetav}\quad&  2\big\lVert{\sum_{m\in\mathcal{M}}\zeta_m\av_m^{(i)}}\big\rVert_2+2\big\lVert{\sum_{m\in\mathcal{M}}\zeta_m\bv_m^{(i)}}\big\rVert_1-\sum_{m\in\mathcal{M}}\zeta_mc_m^{(i)}\\
		\text{s.t. }\quad& \zetav\geq {\bf 0}, \sum_{m\in\mathcal{M}}K_m^2\zeta_m=1.
	\end{align}
\end{subequations}
Note that we cannot guarantee strong duality since the prime problem in \eqref{eq32} is nonconvex. Therefore, the solution in \eqref{eq33} is sub-optimal in general.
}

\begin{algorithm}[!t]
	\caption{SCA-Based Optimization Algorithm on \eqref{eq29}}
	\label{alg1}
	\begin{algorithmic}[1]
		\STATE\textbf{Input:}  $\xv$; $\tau$; $I_{\max}$; and $\e$.\\
		\STATE	\textbf{Initialization}: $\thetav^{(1)}$ and $\fv^{(1)}$ .\\
		\STATE	Compute $\text{obj}^{(1)}=\min_{m:m\in\mathcal{M}} -\abs{(\fv^{(1)})^H\hv_m(\thetav^{(1)})}/K_m^2$;\\
		\STATE \textbf{for} $i=1,2,\cdots,I_{\text{max}}$\\
		\STATE~~Compute $\zetav$ by solving \eqref{eq34};\\
		\STATE~~Compute $\fv^{(i+1)}$ and $\thetav^{(i+1)}$ by \eqref{eq33};\\
		\STATE~~Update $\text{obj}^{(i+1)}=\min_{m\in\mathcal{M}} -\abs{(\fv^{(i+1)})^H\hv_m(\thetav^{(i+1)})}/K_m^2$;\\
		\STATE~~\textbf{if} {$\frac{\abs{\text{obj}^{(i+1)}-\text{obj}^{(i)}}}{\abs{\text{obj}^{(i+1)}}}\leq \e$}, \textbf{early stop};\\
		\STATE\textbf{end for} \\
		\STATE	\textbf{Output:} {$\fv^{(i+1)}$ and $\thetav^{(i+1)}$.}
	\end{algorithmic}
\end{algorithm}
We summarize the algorithm of optimizing $\fv$ and $\thetav$ in Algorithm \ref{alg1}, where the convex problem \eqref{eq34} is solved by a standard convex optimization solver such as SciPy \cite{scipy}. Besides, the proposed algorithm terminates when either the maximum number $I_{\text{max}}$ of iterations is reached or when the change of the objective value between two consecutive iterations is less than a predetermined small number $\e$.
\subsection{Optimizing Device Selection}\label{sec4b}
In this subsection, we propose an iterative device selection algorithm for optimizing $\xv$ based on Gibbs sampling \cite{PowerControl_Qian}. 
Specifically, at iteration $j=1,2,\cdots,J_{\text{max}}$, given the result in the last iteration $\xv_{j-1}$, we define $\widetilde \xv_{j-1}^{(m)}$, $1\leq m\leq M$, such that $\widetilde x_{j-1,m}^{(m)} \neq x_{j-1,m}$ and $\widetilde x_{j-1,m^\prime}^{(m)} = x_{j-1,m^\prime}$ for $ \forall {m^\prime}\neq m$. In other words,  $\widetilde \xv_{j-1}^{(m)}$ differs from $\xv_{j-1}$ only at the $m$-th entry. 
For notational convenience, we define $\tilde \xv_{j-1}^{(0)}=\xv_{j-1}$. Then, we  sample $\xv_j$ from $\mathcal{F}_{j-1}\triangleq\{\widetilde\xv_{j-1}^{(m)}:0\leq m \leq M\}$ according to the following distribution
\begin{align}
	\label{eq35}
	\Lambda_j(\xv_j)=\frac{\exp\left(-J(\xv_j)/\beta_j \right) }{\sum_{\xv\in\mathcal{F}_{j-1}}\exp\left(-J(\xv)/\beta_j \right)},
\end{align}
where $J(\xv)=\min_{\fv,\thetav}d(\xv,\fv,\thetav)$ is the objective value given $\xv$ that is obtained by Algorithm \ref{alg1}; and $\beta_j>0$ denotes the ``temperature" parameter. Here, we employ a slowly decreasing ``cooling schedule" \cite{Annealing} for $\{\beta_j\}$ to accelerate the convergence by $\beta_{j}=\rho\beta_{j-1}$ for some $0<\rho<1$. 
\begin{algorithm}[!t]\label{A1}
	\caption{Gibbs-Sampling Based Algorithm on \eqref{eq28}}
	\label{alg2}
	\begin{algorithmic}[1]
		\STATE		\textbf{Input:} $\{K_m\}$,  $\{\hv_{DP,m}\}$, $\{\Gv_m\}$, $\sigma^2_n$, $P_0$, $\beta_0$, $\rho$, $J_{\max}$.\\
		\STATE	\textbf{Initialization}: $\xv_0=[1,1,...,1]$.\\
		\STATE \textbf{for} $j=1,2,\cdots,J_{\text{max}}$\\
		\STATE~~Generate $\mathcal{F}_{j-1}$;\\
		\STATE~~\textbf{for} every $\tilde \xv_{j-1}^{(m)}$ in $\mathcal{F}_{j-1}$ \\
		\STATE~~~~Compute $(\fv_{j-1}^{(m)},\thetav_{j-1}^{(m)})=\argmin_{\fv,\thetav}d(\tilde \xv_{j-1}^{(m)},\fv,\thetav)$ by Algorithm \ref{alg1};\\
		\STATE~~\textbf{end for} \\
		\STATE~~Sample $\xv_j$ according to \eqref{eq35};
		\STATE~~Update $\b_{j}= \rho\b_{j-1}$;
		\STATE\textbf{end for} \\
		\STATE	\textbf{Output:} {$\xv_{J_{\max}}$ together with the corresponding solution to $(\fv,\thetav)$.}
	\end{algorithmic}
\end{algorithm}

The overall algorithm is summarized in Algorithm \ref{alg2}. In order to compute $J(\xv)$ for $\forall \xv\in\mathcal{F}_{j-1}$ in \eqref{eq35}, Algorithm \ref{alg1} is repeatedly invoked to solve \eqref{eq29}. To reduce the computational complexity, we introduce an additional warm-start technique in Algorithm \ref{alg2}. Specifically, we note that, for $\forall 0\leq m\leq M$, the decision vector  $\widetilde{\xv}_{j-1}^{(m)}$ differs with $\widetilde{\xv}_{j}^{(m)}$ at most $2$ entries. Therefore, the corresponding solution $(\fv_{j}^{(m)},\thetav_{j}^{(m)})$ in Line 6 of Algorithm \ref{alg2} is not much different from  $(\fv_{j-1}^{(m)},\thetav_{j-1}^{(m)})$ in most cases. Motivated by this observation, we use $(\fv_{j-1}^{(m)},\thetav_{j-1}^{(m)})$ as an initial point in computing $(\fv_{j}^{(m)},\thetav_{j}^{(m)})$, so that Algorithm \ref{alg1} can early stop with a few iterations.
\subsection{Computational Complexity}
When \eqref{eq34} is solved by the existing convex optimization solver, the interior point method is considered, whose worst-case complexity is bounded by $O(\abs{\mathcal{M}}^3)$ for a given $\mathcal{M}$. As a result, the complexity of Algorithm \ref{alg1} is upper bounded by $O(I_{\text{max}}M^3)$. Since Algorithm \ref{alg2} invokes Algorithm \ref{alg1} $J_{\text{max}}M$ times, the overall complexity of the proposed algorithm is bounded by $O(J_{\text{max}}I_{\text{max}}M^4)$. 

In contrast, the complexity of the alternating optimization algorithm in \cite{FL_RIS2} is $O(I_{\text{max}} (N^6+L^6))$, which is much larger than that of Algorithm \ref{alg1}. As another baseline, the FL optimization algorithm in \cite{FL_1} jointly optimizes the device selection decision $\mathcal{M}$ and the receiver beamforming vector  $\fv$ without considering the RIS. Its complexity  is $O(M(N^6+(N^2+M)^3))$, which has a larger order than that of Algorithm \ref{alg2}.

\section{Numerical Results}\label{sec6}
In this section, we conduct numerical experiments to examine the proposed optimization algorithm and  compare the proposed algorithm with the state-of-the-art approaches. The simulation setup is given in Section \ref{sec6a}. The simulation codes of our algorithm are available at https://github.com/liuhang1994/RIS-FL.
\begin{figure}[!t]
	\centering
	\includegraphics[width=3.8in]{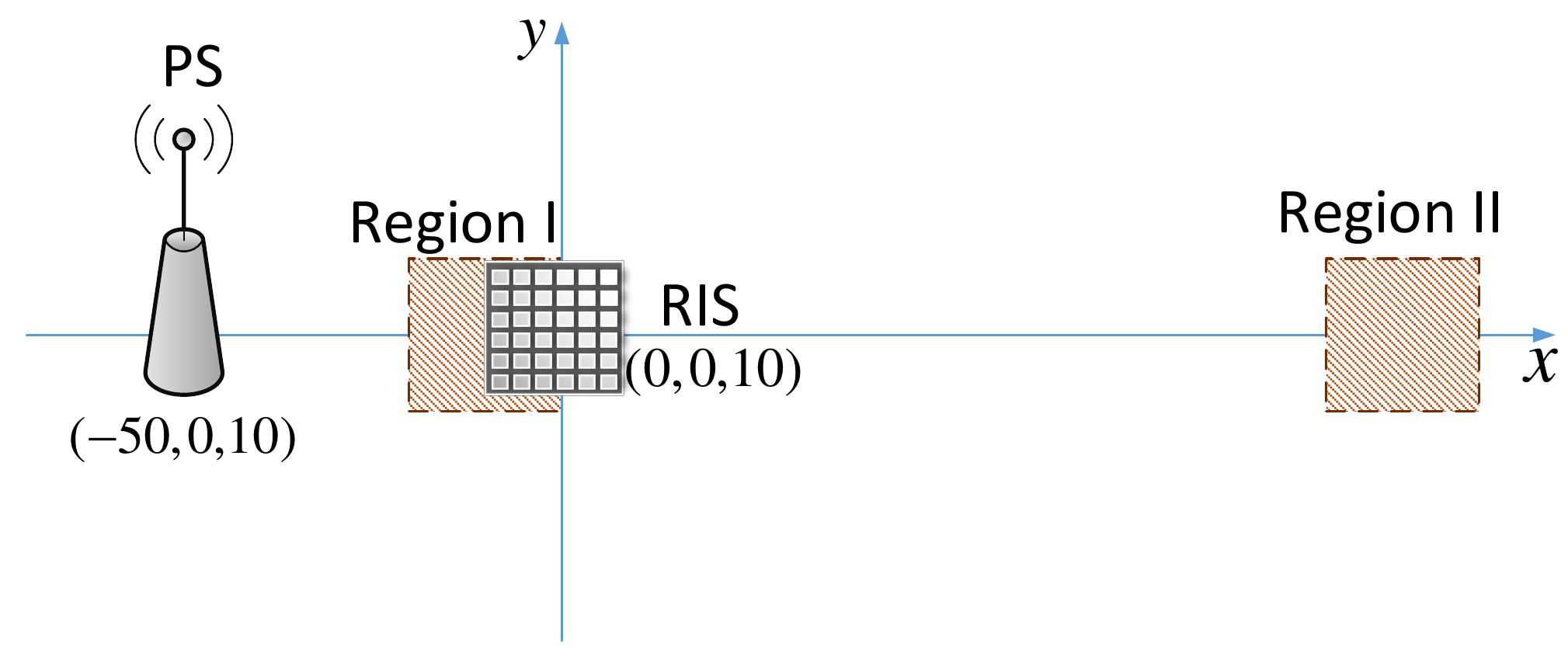}
	\caption{Simulation setup of the RIS-assisted system (top view).}\label{Fig3}
\end{figure}
\subsection{Simulation Setup}\label{sec6a}
\begin{table*}[t]
	\centering
	\caption{System Parameters}\label{table1}
	\medskip
	\begin{tabular}{|l|l||l|l||l|l|}
		\hline
		Parameter&Value&Parameter&Value&Parameter&Value\\
		\hline
		$N$&$5$&$I_{\text{max}}$&$100$&$G_{PS}$&$5 $ dBi\\
		\hline
		$L$&$40$&$\e$&$0.01$&$G_{D}$  &$0$ dBi\\
		\hline
		$M$&$40$&$\tau$&$1$&$G_{RIS}$ &$5 $ dBi \\
		\hline
		$P_0$&$-10$ dB&	$\rho$&$0.9$&$f_c$ &$915 $ MHz\\
		\hline
		$\sigma_n^2$&$-100$ dB&$\b_0$&$1$&$d_x$ &$(3*10^7 m/s)/f_c$\\
		\hline
		$J_{\text{max}}$&$50$&$PL$&$3.76$&$d_y$ &$(3*10^7 m/s)/f_c$\\

		\hline
	\end{tabular}
\end{table*}

We consider a three-dimensional (3D) Cartesian coordinate system shown in Fig. \ref{Fig3}. The PS is placed at $(-50,0,10)$ and the RIS is placed at $(0,0,10)$. The locations of the edge devices will be specified later. The channel coefficients are given by the small-scale fading coefficients multiplied by the square root of the path loss. The small-scale fading coefficients follow the standard independent and identically distributed (\iid) Gaussian distribution in Sections \ref{sec6c}--\ref{sec6f}, and follow the Rician channel model in Section \ref{sec6g}. The free-space path loss model is adopted for the device-PS direct channels as
	$G_{PS}G_{D}\left( \frac{3*10^8 m/s}{4\pi f_cd_{DP}}\right) ^{PL}$,
where $G_{PS}$ (or $G_{D}$) is the antenna gain at the PS (or the devices); $f_c$ is the carrier frequency; $PL$ is the path loss exponent; and $d_{DP}$ is the distance between the device and the PS.
Meanwhile, the path loss of the cascaded device-RIS-PS channel is modelled as 
	$G_{PS}G_{D}G_{RIS}\frac{L^2d_xd_y((3*10^8 m/s)/f_c)^2}{64\pi^3d_{RP}^2d_{DR}^2}$ \cite[Proposition 1]{tang2019wireless}.
Here, $G_{RIS}$ is the antenna gain at the RIS; 
$d_x$ (or $d_y$) is the horizontal (or vertical) size of a single RIS element; and $d_{RP}$ (or $d_{DR}$) is the distance between the RIS and the PS (or the device and the RIS). {Unless otherwise specified, the values of the system parameters are given in Table \ref{table1}.}


We simulate the image classification task on the Fashion-MNIST dataset \cite{Xiao2017}. Specifically, we train a convolutional neural network with two $5\times5$ convolution layers (each with $2\times 2$ max pooling) followed by a  batch normalization layer, a fully connected layer with $50$ units, a ReLu activation layer, and a softmax output layer (total number of parameters $D=21,921$). The loss function is the cross-entropy loss. The local training data are \iid drawn from $50,000$ images with $\{K_m\}$ specified later, and the test dataset contains $10^4$ different images. The FL learning performance is evaluated by the test accuracy defined as $\frac{\text{the number of correctly classified test images}}{10,000}\in[0,1]$.

Finally, we consider the following two settings on the numbers of local training samples $\{K_m\}$ and the locations of the devices:
\begin{description}
	\item[Setting 1] Concentrated devices with equal data sizes : The $M$ devices are uniformly allocated in a rectangular region (\ie $\text{Region I}\triangleq\{(x,y,0):-20\leq x\leq 0,-10\leq y\leq 10\}$). Moreover, every $K_m$ is set to $750$.
	
	\item[Setting 2] Two-cluster devices with unequal data sizes: The $M$ devices are randomly allocated in two rectangular regions: half of the $M$ devices are randomly located in the near region (\ie $\text{Region I}$); and the other half are located in the far region (\ie $\text{Region II}\triangleq\{(x,y,0):100\leq x\leq 120,-10\leq y\leq 10\}$). Meanwhile, the devices have different numbers of the training samples. We randomly select half devices and draw the corresponding values of $\{K_m\}$ uniformly from $[1000,2000]$. The values of $\{K_m\}$ for the other half devices are uniformly drawn from $[100,200]$. 
\end{description}
Intuitively, the first setting allocates the same number of samples to the devices and locates all the devices in a small region so that the channel path loss coefficients are similar among the devices. In other words, all the devices behave similarly without any strong straggler issue. 
Otherwise, the second setting models the channel heterogeneity (\ie the near-far effect) and the data heterogeneity (\ie the different data sizes) by separating the devices based on their locations and data sizes. In this case, the straggler issue becomes severe, and device selection is important.

\subsection{Simulations on Batch Gradient Descent}\label{sec6c}
In this section, we examine the performance of the proposed algorithm for the image classification task described in Section \ref{sec6a}.  We study the FL framework in Section \ref{sec2a} with the gradient computation in \eqref{eq03}. The learning rate $\lambda$ is set to $0.01$.
We consider the following ideal benchmark, which characterizes the best possible FL performance in the error-free case.
\begin{itemize}
	\item Error-free channel: Suppose the channels are noiseless  (\ie $\sigma^2_n=0$). All the devices are selected to participate in the model aggregation. The PS receives all the local gradients in a noiseless fashion and updates the global model by \eqref{eq04} with $\hat \rv=\rv$.
\end{itemize}
Moreover, the following baselines are used for comparison: 
\begin{itemize}
	\item DC-based alternating optimization without device selection \cite{FL_RIS2}: Suppose that all the devices are selected, \ie $\mathcal{M}=\{1,2,\cdots,M\}$. The variables $\fv$ and $\thetav$ are alternately optimized by DC programming in order to minimize the communication MSE.
	\item DC-based optimization without RIS \cite{FL_1}: Suppose that the RIS is not considered, \ie $\thetav=\bf 0$. The device selection and the receiver beamforming are jointly optimized by DC programming such that the number of active devices $\abs{\mathcal{M}}$ is maximized while the communication MSE is below some threshold $\gamma$.\footnote{From numerical experiments, we find that $\gamma= 15$ dB yields the best performance for the algorithm in \cite{FL_1} in most cases. Therefore, we set $\gamma=15$ dB in the implementation of this algorithm.}
	\item Receiver beamforming by differential geometry programming \cite{FL_DG}: Suppose $\thetav=\bf 0$ and $\mathcal{M}=\{1,2,\cdots,M\}$. In order to minimize the communication MSE, the receiver beamforming vector is optimized via  differential geometry programming.
\end{itemize}
 \begin{figure}[!t]
	
	\begin{minipage}[t]{0.5\linewidth}
		\centering
		\includegraphics[width=3 in]{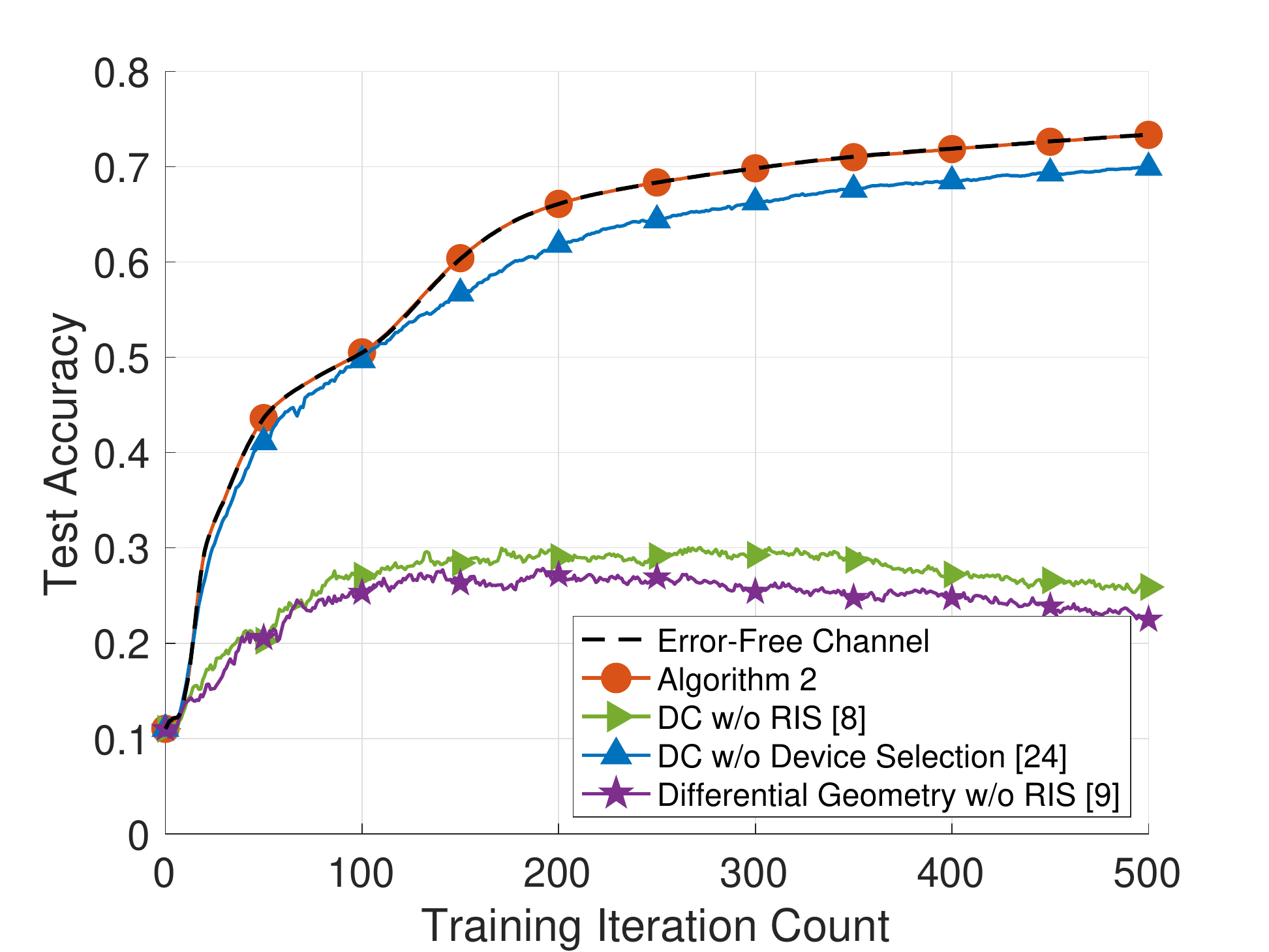}
		\caption{Test accuracy of the proposed algori-\protect\\thm with $M=L=40$, concentrated devices,\protect\\ equal data sizes (Setting 1).}
		\label{set1_ite}
	\end{minipage}
	\begin{minipage}[t]{0.5\linewidth}
		\centering
		\includegraphics[width=3 in]{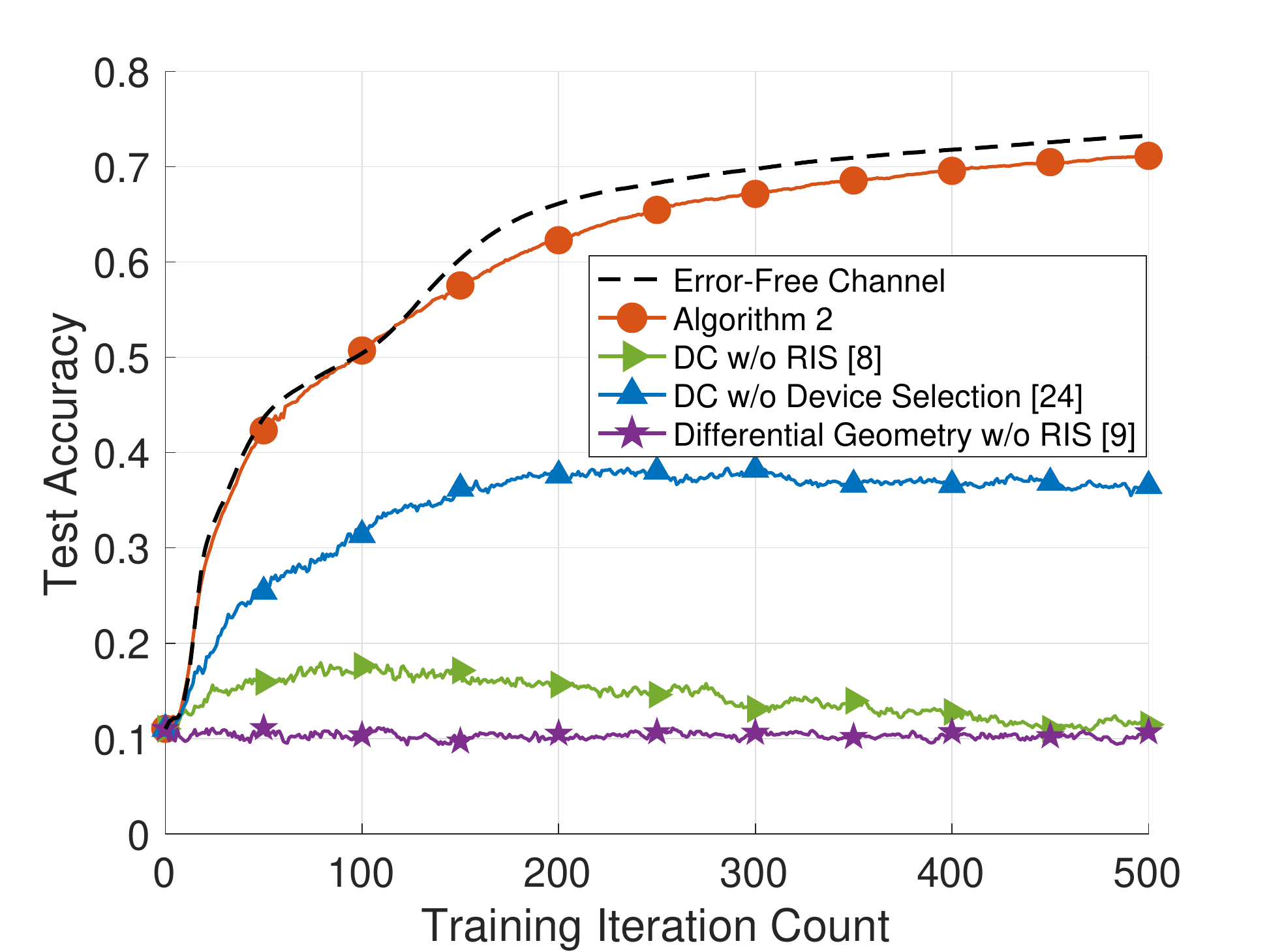}
		\caption{Test accuracy with two-cluster devices and unequal data sizes (Setting 2).}
		\label{set2_ite}
	\end{minipage}%
\end{figure}

We plot the test accuracy of the proposed algorithm in $T=500$ training rounds with concentrated devices and equal data sizes (\ie Setting 1) in Fig. \ref{set1_ite}. We set $N=5$ and $M=L=40$. The results are averaged over $50$ Monte Carlo trials. As discussed in the previous subsection, this setting has no strong straggler issue, and thus the effect of device selection becomes marginal. Therefore, we see that the proposed method achieves similar accuracy to the method in \cite{FL_RIS2} in which device selection is not considered. Moreover, compared with the FL algorithms without RISs (\ie the baselines in  \cite{FL_1,FL_DG}) that suffer from large aggregation errors, the RIS-enabled FL algorithms including Algorithm 2 and the baseline in \cite{FL_RIS2} overcome the unfavorable channel conditions by configuring the RIS phase shifts. 

In Fig. \ref{set2_ite}, we plot the test accuracy with two-cluster device locations and unequal data sizes (\ie Setting 2). In this setting, the devices have varying channel conditions and varying data sizes. We see from Fig.  \ref{set2_ite} that the straggler issue heavily affects the performance of the RIS-enabled FL without device selection (\ie the method in \cite{FL_RIS2}). In contrast, thanks to the efficient device selection, the proposed algorithm still maintains the near-optimal accuracy, verifying the robustness of the proposed method.

\begin{figure}[!t]
	\begin{minipage}[t]{0.5\linewidth}
		\centering
		\includegraphics[width=3 in]{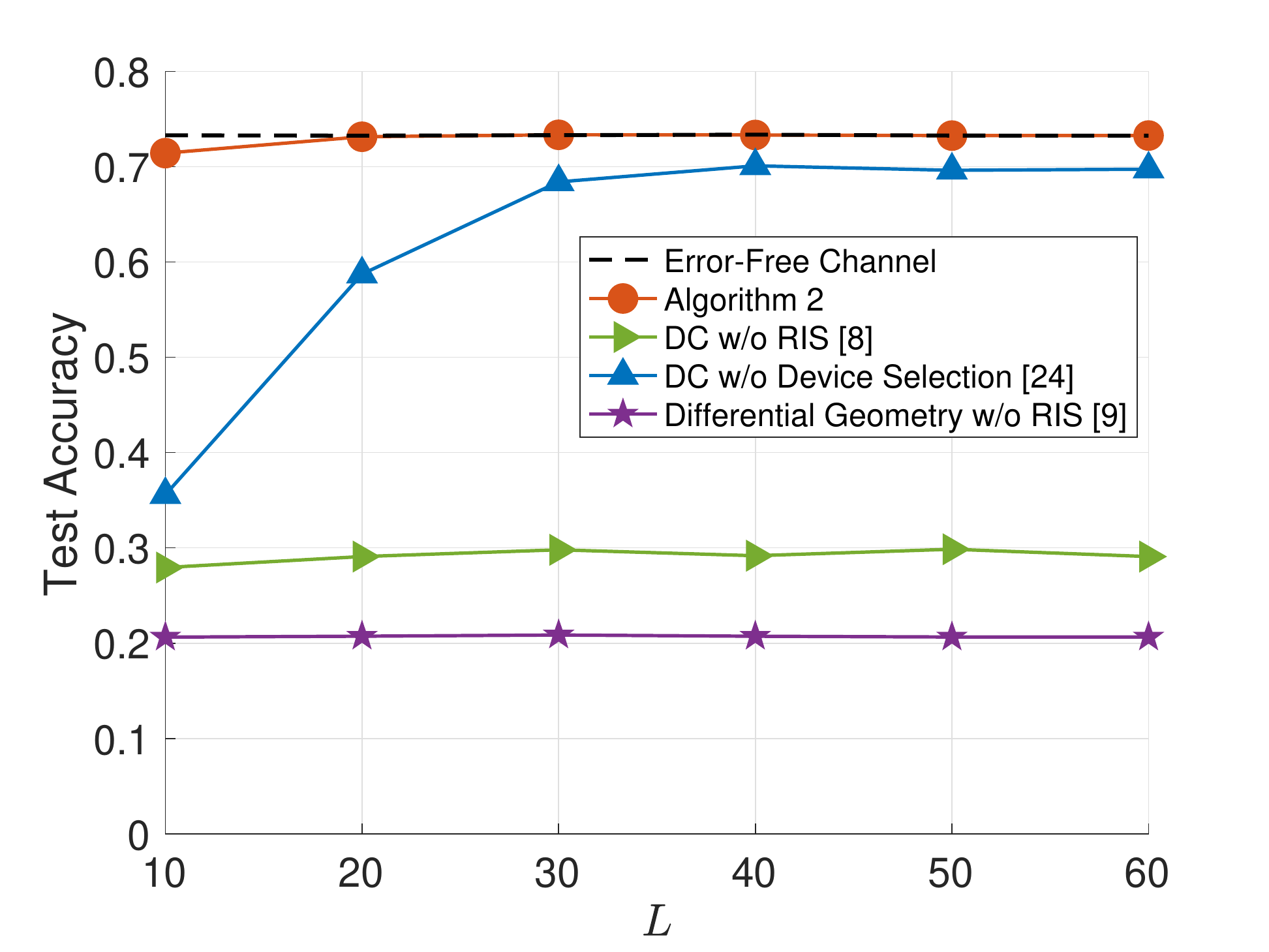}
		\caption{Test accuracy versus $L$ with Setting 1.}
		\label{set1_L}
	\end{minipage}
	\begin{minipage}[t]{0.5\linewidth}
		\centering
		\includegraphics[width=3 in]{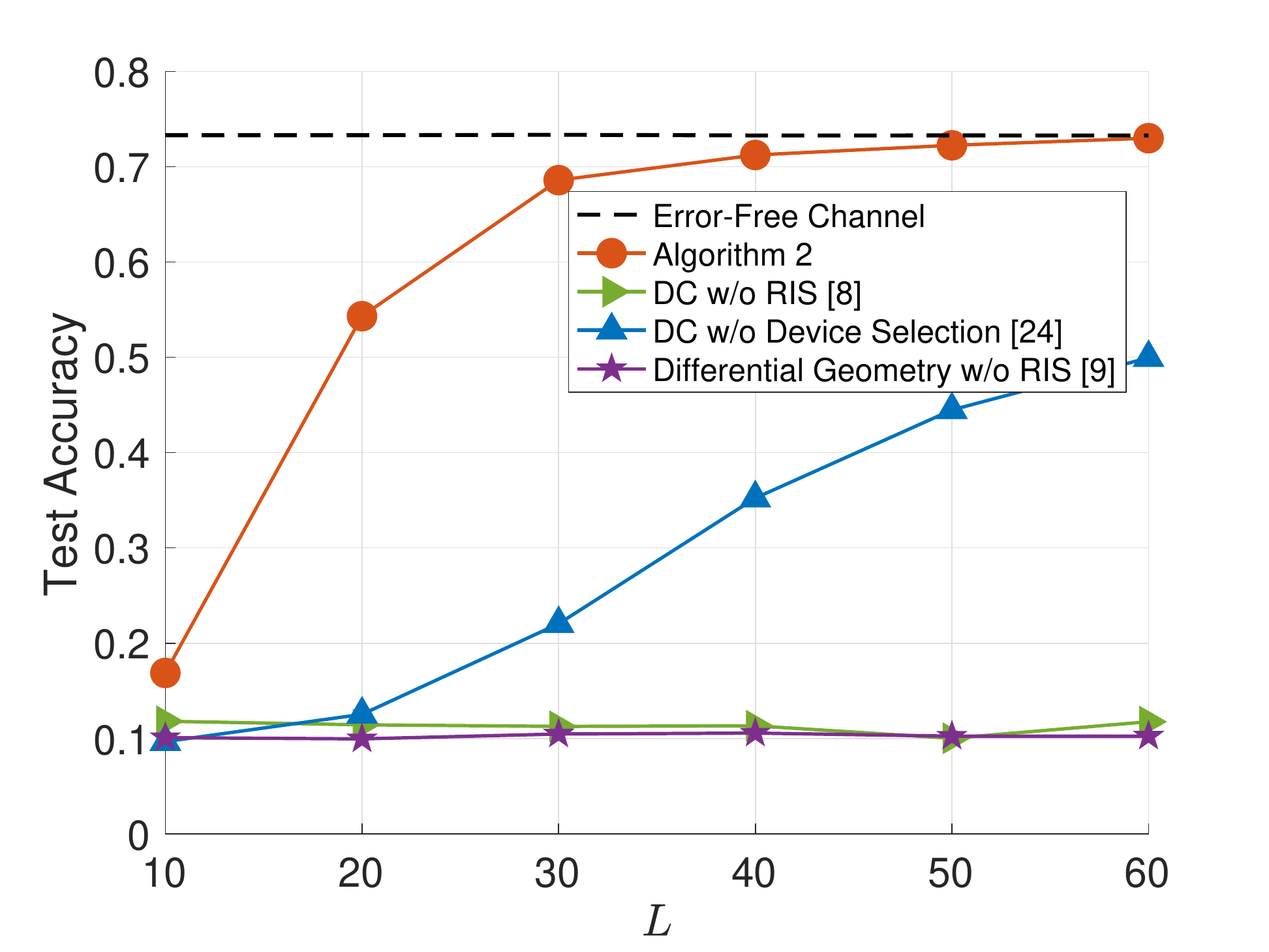}
		\caption{Test accuracy versus $L$ with Setting 2.}
		\label{set2_L}
	\end{minipage}%
\end{figure}
In Figs. \ref{set1_L} and \ref{set2_L}, we investigate the test accuracy versus the RIS size $L$ under Settings 1 and 2, respectively. Comparing the proposed method with the baseline in \cite{FL_RIS2}, we see that both methods achieve considerable performance improvement with a large $L$ under Setting 1; see Fig. \ref{set1_L}. However, in the case with strong straggler issues, our method outperforms the one in \cite{FL_RIS2} as shown in Fig. \ref{set2_L}. This is because the device selection avoids the aggregation error explosion by excluding the weak devices. Meanwhile, the baselines in \cite{FL_1,FL_DG} do not consider RIS phase shifts, and hence their results are invariant to the RIS size $L$.
\subsection{Simulations on Mini-Batch Gradient Descent}\label{sec6d}
In this subsection, we consider a different FL framework, where the local model update  in \eqref{eq03} is replaced by multiple mini-batch gradient descent updates with the learning rate $\lambda=0.005$. Specifically, we randomly partition the local dateset $\mathcal{D}_m$ into $\lceil {K_m/B}\rceil$ disjoint batches $\mathcal{D}_m^{(1)},\mathcal{D}_m^{(1)},\dots,\mathcal{D}_m^{(\lceil{K_m/B}\rceil)}$, each having $B$ training samples\footnote{If $K_m/B$ is not an integer, the last batch contains $K_m-B(\lceil{K_m/B}\rceil-1)$ samples.}, where $\lceil \cdot\rceil$ is the ceiling function. After receiving the current global model $\wv_t$, the $m$-th device performs gradient descent with respect to these batches sequentially by
\begin{align}
	&\wv_{m,t}^{(1)}=\wv_t-\lambda \nabla F_m(\mathcal{D}_m^{(1)};\wv_t), \wv_{m,t}^{(2)}=\wv_{m,t}^{(1)}-\lambda \nabla F_m(\mathcal{D}_m^{(2)};\wv_{m,t}^{(1)}),\nonumber\\
	&\cdots\cdots\nonumber\\
	&\wv_{m,t}^{(\lceil{K_m/B}\rceil)}=\wv_{m,t}^{(\lceil{K_m/B}\rceil-1)}-\lambda \nabla F_m(\mathcal{D}_m^{(\lceil{K_m/B}\rceil)};\wv_{m,t}^{(\lceil{K_m/B}\rceil-1)}).
\end{align}
The corresponding gradient vector in \eqref{eq03} is replaced by the model change as $\gv_{m,t}=(\wv_{m,t}^{(\lceil{K_m/B}\rceil)}-\wv_t)/\lambda$ and the corresponding global model aggregation process remains the same as in \eqref{eq04}.

\begin{figure}[!t]	
	\begin{minipage}[t]{0.5\linewidth}
		\centering
		\includegraphics[width=3 in]{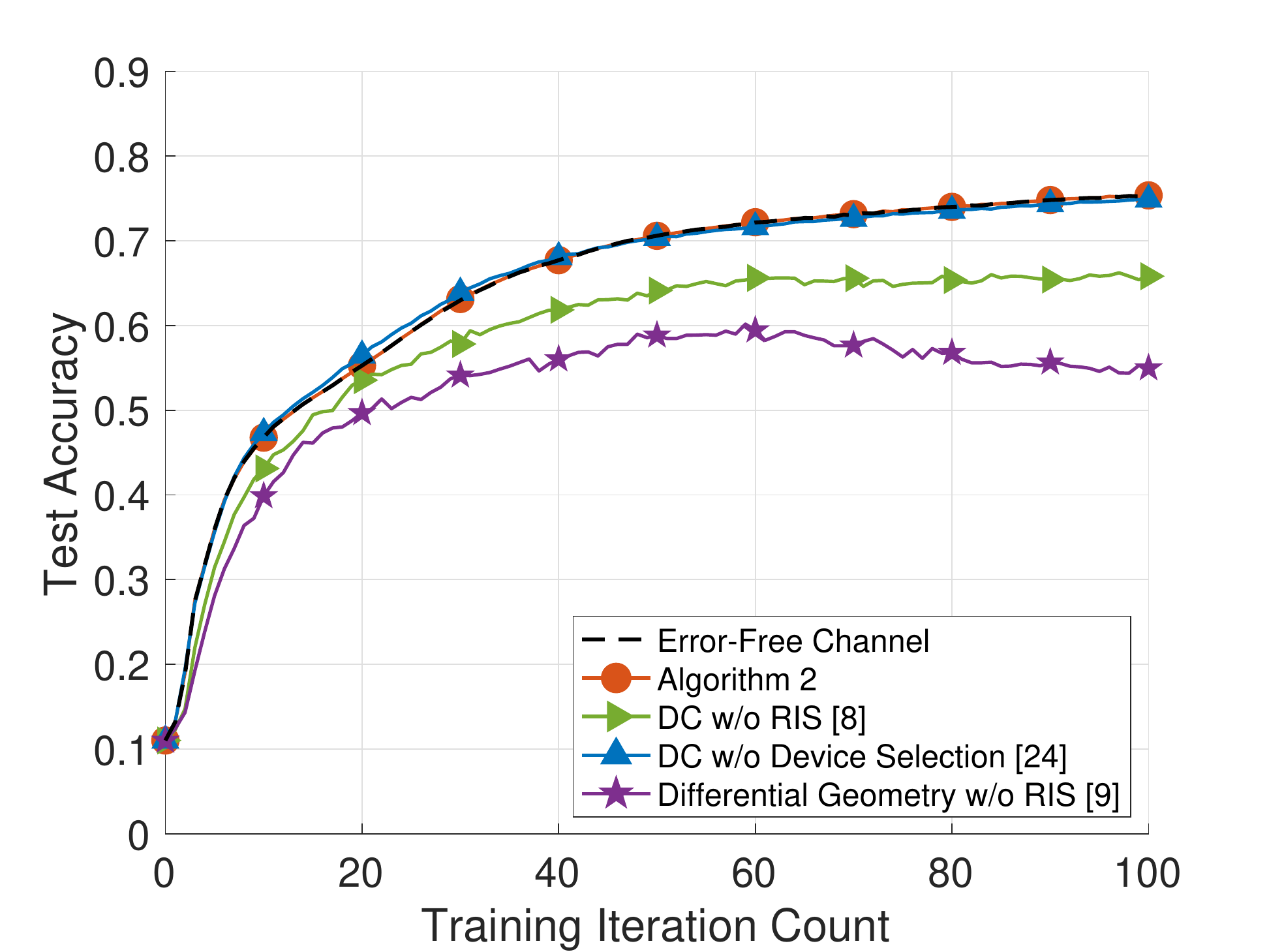}
		\caption{Mini-batch gradient descent test accu-\protect\\racy with concentrated devices and equal data sizes (Setting 1).}
		\label{set1_ite_MB}
	\end{minipage}
	\begin{minipage}[t]{0.5\linewidth}
		\centering
		\includegraphics[width=3 in]{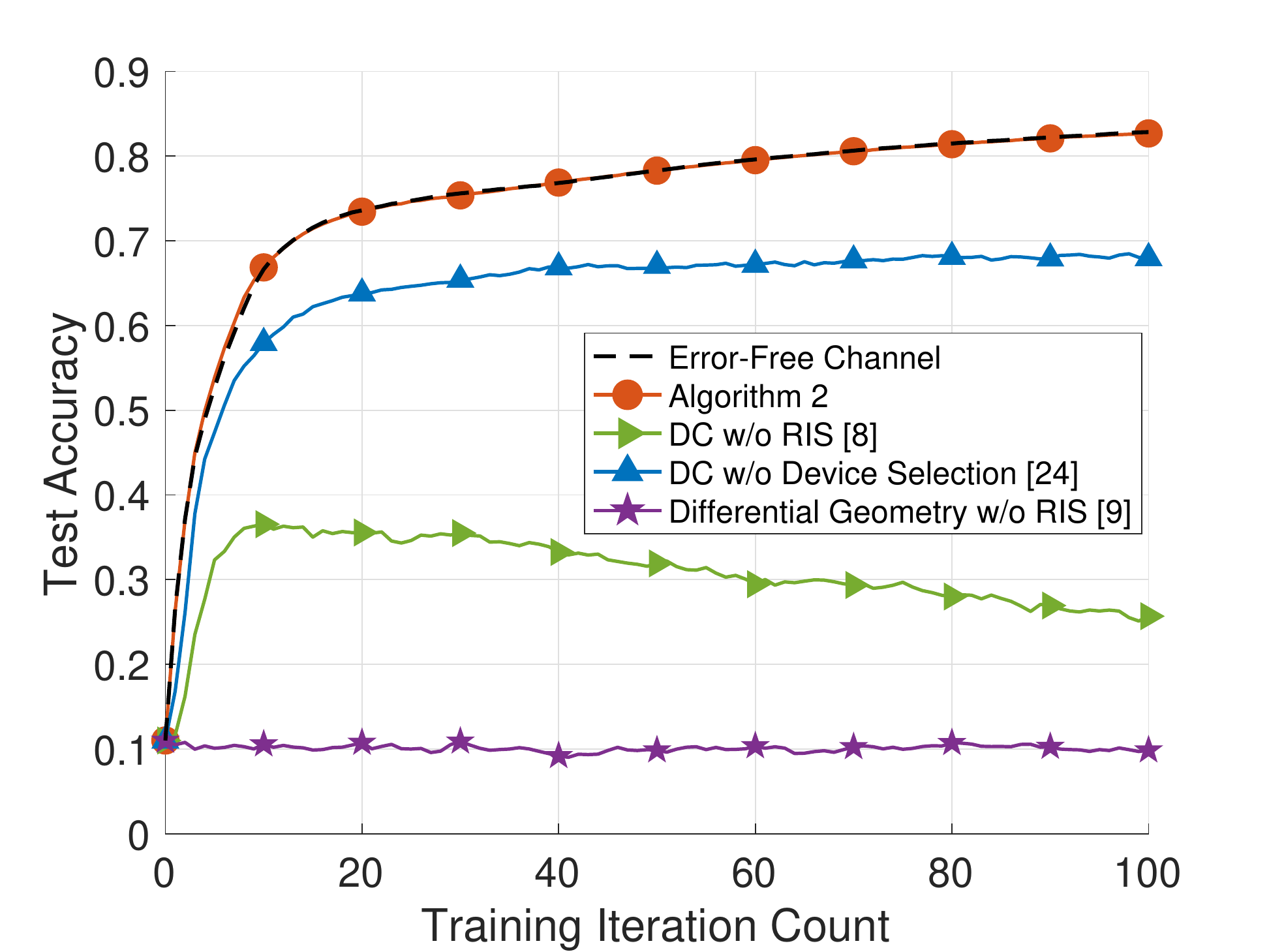}
		\caption{Mini-batch gradient descent test accuracy with two-cluster devices and unequal data sizes (Setting 2).}
		\label{set2_ite_MB}
	\end{minipage}%
\end{figure}
In Figs. \ref{set1_ite_MB} and \ref{set2_ite_MB}, we plot the test accuracy versus the training round $T$ with Settings 1 and 2, respectively, where we set $B=128$, $N=5$, and $M=L=40$. We see that all the simulated FL algorithms converged with relatively high prediction accuracy under Setting 1. This is because there is no strong straggler issue in this setting. Similarly to the batch gradient descent case, device selection is critical in Setting 2, as it overcomes the straggler issue by striking a good balance between data exploitation and communication error mitigation. In this case, our algorithm outperforms all the baselines; see Fig.  \ref{set2_ite_MB}.
\begin{figure}[!t]
	\begin{minipage}[t]{0.5\linewidth}
		\centering
		\includegraphics[width=3 in]{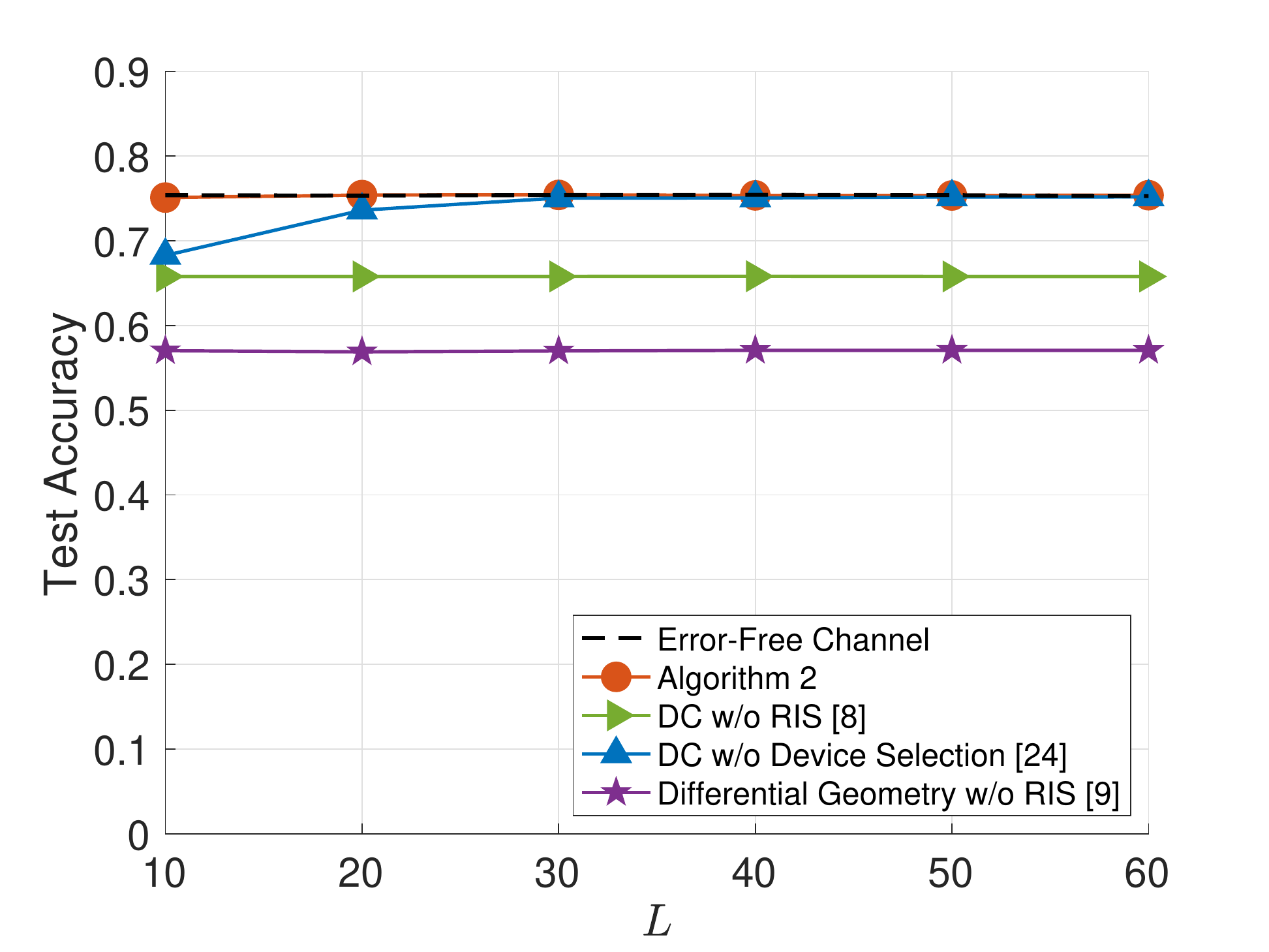}
		\caption{Mini-batch gradient descent test  accu-\protect\\racy versus $L$ with Setting 1.}
		\label{set1_L_MB}
	\end{minipage}
	\begin{minipage}[t]{0.5\linewidth}
		\centering
		\includegraphics[width=3 in]{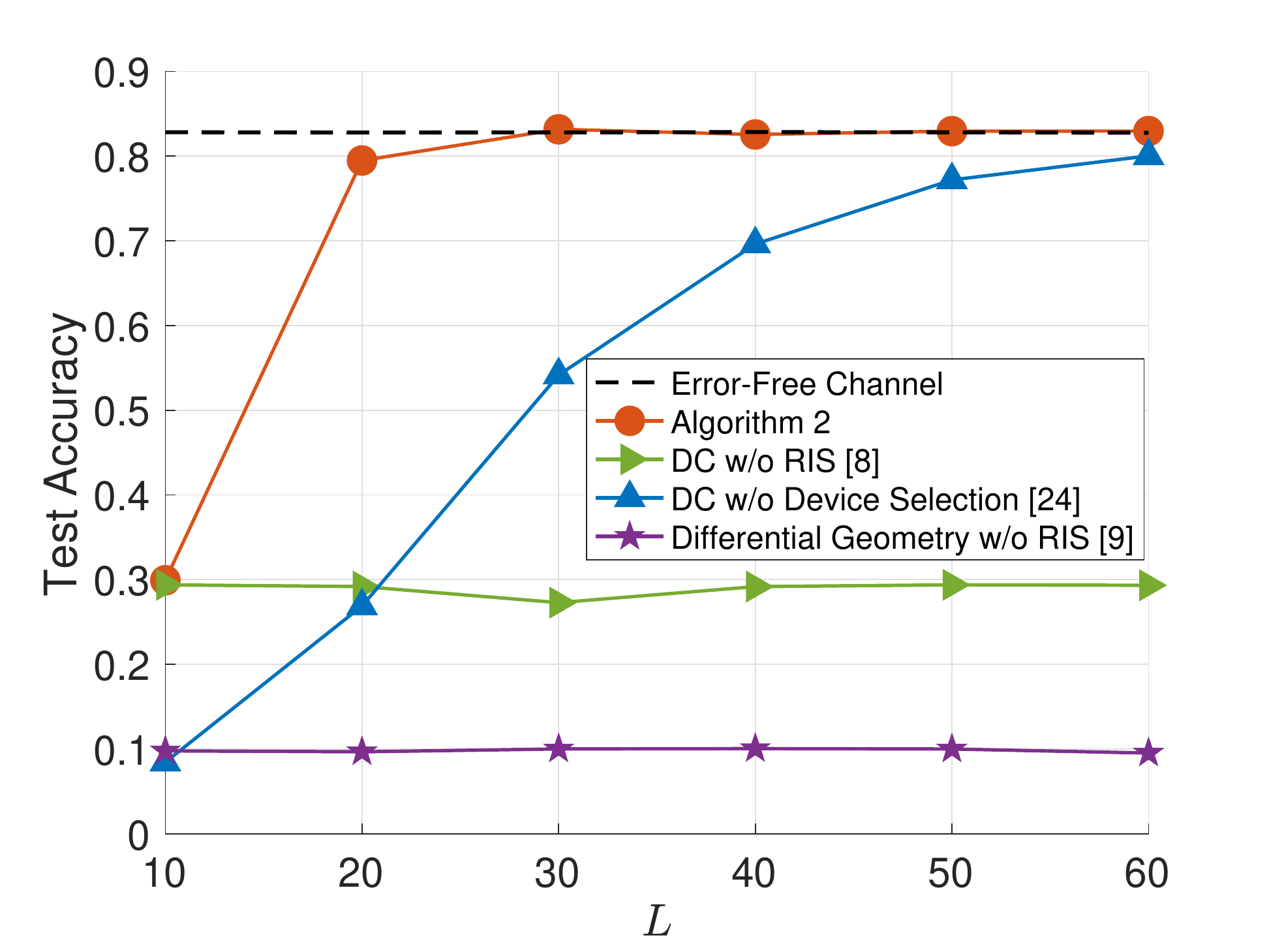}
		\caption{Mini-batch gradient descent test accuracy versus $L$ with Setting 2.}
		\label{set2_L_MB}
	\end{minipage}%
\end{figure}

Finally, we plot the FL performance versus the RIS size $L$ in Figs. \ref{set1_L_MB} and \ref{set2_L_MB} with Settings 1 and 2, respectively. We find that the proposed method requires a smaller $L$ to achieve near-optimal accuracy  than the baseline in \cite{FL_RIS2}. Specifically, the proposed method requires $L\geq 10$ and $L\geq 30$ in Figs. \ref{set1_L_MB} and \ref{set2_L_MB}, respectively, while the requirements of the baseline in \cite{FL_RIS2} are $L\geq 30$ and $L\geq 60$. This observation verifies the superiority of the proposed method, especially when the stragglers exist.
{\subsection{Simulations on Discrete RIS Phase Shifts}\label{sec6f}
Recent studies in \cite{wu2020intelligent,LIS_RZhang_discrete} show that the continuous phase shift model $\theta_l\in\{\theta\in\Complex: |\theta_l|^2=1\},\forall l,$ may incur high implementation cost in practice. 
In this subsection, we assume that 
the RIS phase shifts can take only a finite number of discrete values by following \cite{LIS_RZhang_discrete}. Let $b$ denote the number of bits of each RIS phase shift. The RIS phase shift vector $\thetav=[\theta_1,\cdots,\theta_L]\in\Complex^{L\times 1}$ is given by
\begin{align}\label{eq07}
	\theta_l\in\Theta_b\triangleq \{\theta=e^{j\frac{2\pi i}{2^b}}:0\leq i\leq 2^b-1\},\forall l,
\end{align}
where $\Theta_b$ is the discrete feasible set that each $\theta_l$ belongs to. Note that when $b\to\infty$, $\Theta_b$ becomes a continuous set, \ie $\Theta_b=\{\theta:|\theta|^2=1\}$, which is used in the previous simulations. 
	\begin{figure}[!t]	
	\begin{minipage}[t]{0.5\linewidth}
		\centering
		\includegraphics[width=3 in]{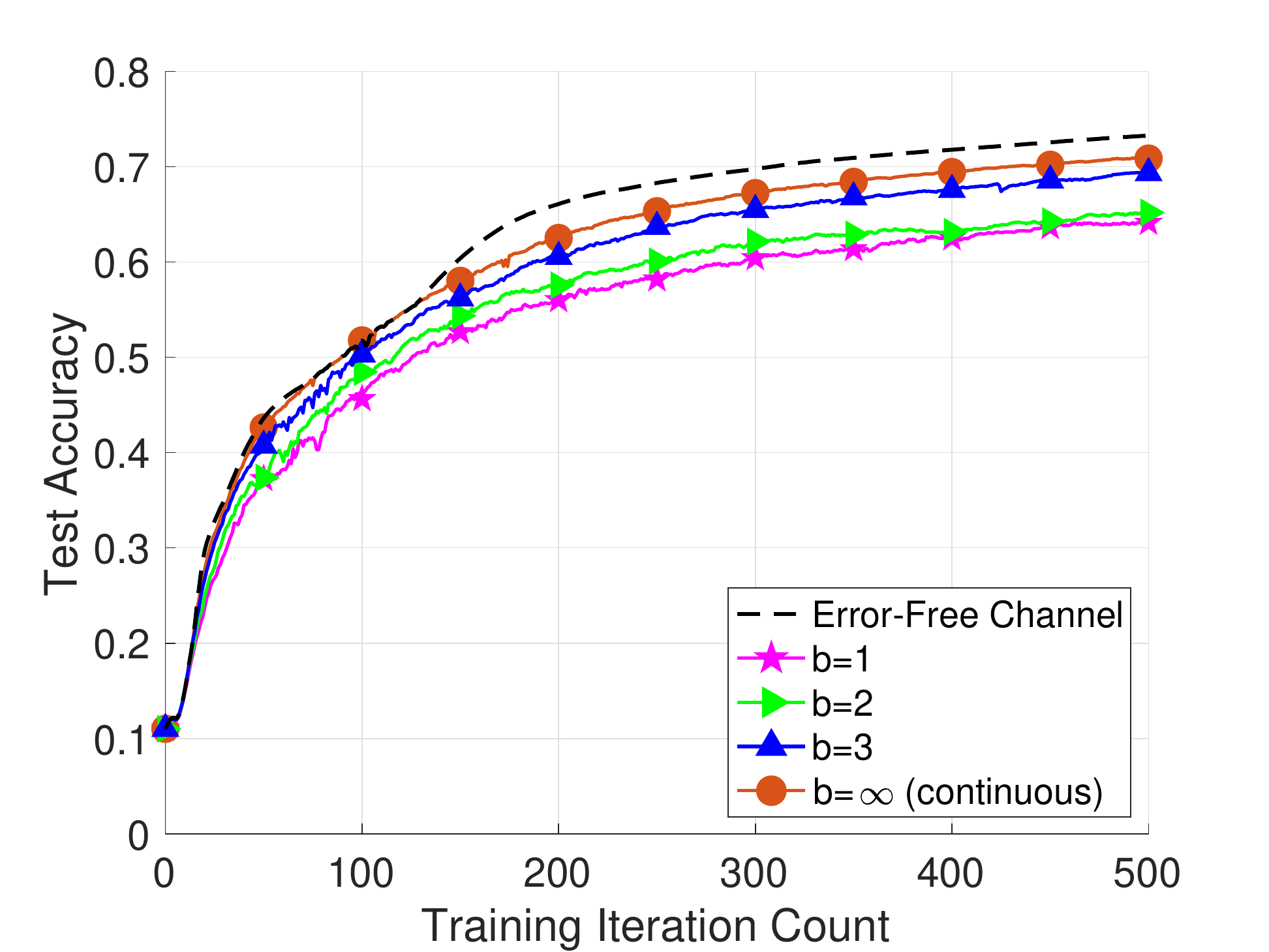}
		\caption{FL accuracy of the proposed algorithm with Setting 2 and discrete phase shift levels. }
		\label{dis}
	\end{minipage}
	\begin{minipage}[t]{0.5\linewidth}
		\centering
		\includegraphics[width=3 in]{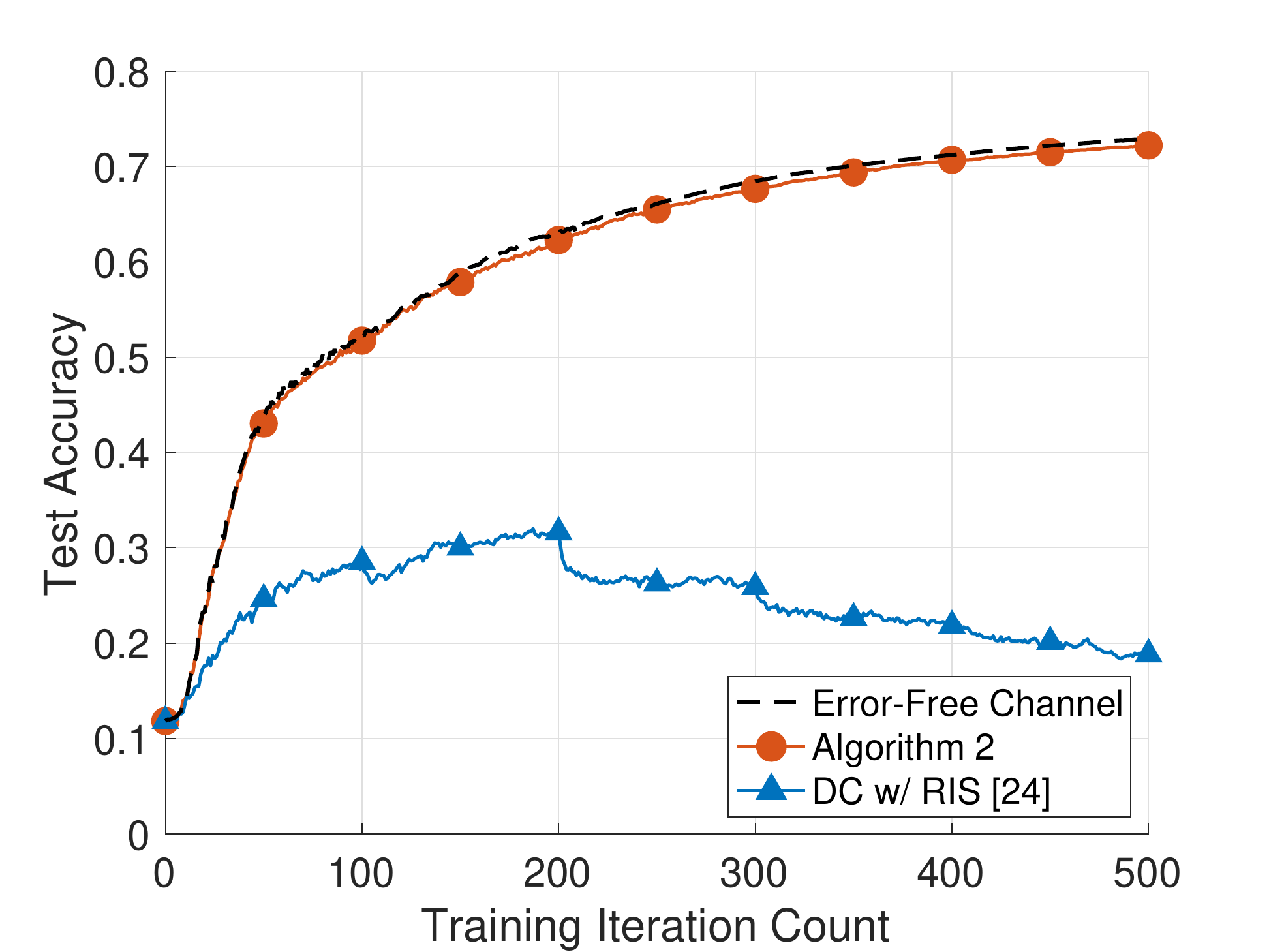}
		\caption{Learning accuracy of the proposed algorithm under time-varying Rician fading channels.}
		\label{vary}
	\end{minipage}%
\end{figure}

Here, we accommodate the proposed algorithm to the case with discrete RIS phase shifts by projecting the solution to \eqref{eq07}. Specifically, we first treat $\thetav$ as if $b=\infty$, \ie 
$\thetav$ can take continuous values. Then, we use Algorithm \ref{alg2} to optimize $\thetav$. Finally, we project the solution from Algorithm  \ref{alg2} to the feasible set given in \eqref{eq07}. Fig. \ref{dis} plots the learning accuracy of Algorithm \ref{alg2} under various choices of $b\in\{1,2,3,\infty\}$  with $M=L=40$, two-cluster devices, and unequal data sizes (\ie Setting 2). Except for the model on $\thetav$, other system parameters are the same as in Section \ref{sec6c}. The error-free benchmark is also included for comparison.  We find that the case with $b=\infty$ yields the best performance ($\approx 0.71$) and its accuracy is close to that of the optimal benchmark ($\approx 0.73$). When $b=1$ or $2$, there exists $\approx 0.1$ accuracy degradation compared with the continuous case $b=\infty$. This is because low-bit phase shifts introduce an additional mismatch and thus lead to the performance loss. When $b=3$, this loss becomes neglectable and the learning accuracy is close to the case with continuous phase shifts. We conclude from Fig. \ref{dis} that our algorithm works well with $b\geq 3$. 
\subsection{Simulations on Time-Varying Rician Fading Channels}\label{sec6g}
In this subsection, we simulate a different channel model where the small-scale fading channel coefficients of all the channels vary independently every $100$ training rounds. Therefore, with $T=500$ rounds, the small-scale fading coefficients change $5$ times, and we need to update the solution everytime when the channels change. Note that the optimization algorithm (Algorithm \ref{alg2}) may output different device selection decisions with different channel coefficients, and hence the selected devices are varying during the training process.  Furthermore, since line-of-sight (LoS) paths may exist in the RIS channels, we consider the Rician fading channel model in this subsection. Specifically, the small-scale fading coefficients in the RIS-PS channel are modelled as
\begin{align}
 \Hv_{RP}=\sqrt{\frac{\chi}{1+\chi}} \Hv_{RP}^{\text{LoS}}+\sqrt{\frac{1}{1+\chi}} \Hv_{RP}^{\text{NLoS}},
\end{align}
where $\chi$ is the Rician factor; $\Hv_{RP}^{\text{LoS}}$ is the deterministic LoS component determined by the locations of the RIS and PS; and $\Hv_{RP}^{\text{NLoS}}$ is the non-LoS component with entries following the \iid standard Gaussian distribution. The device-PS and device-RIS channels are generated by following the same procedure. We set the Rician factor to be $3$ dB, $0$, and $0$ for the RIS-PS, the device-PS, and the device-RIS channels, respectively.
 We plot the learning accuracy of the proposed algorithm with time-varying channels, two-cluster devices, and unequal data sizes (\ie Setting 2) in Fig. \ref{vary}, where $M=L=40$ and other system parameters are the same as in Fig. \ref{set2_ite}. We find that our algorithm still achieves an accuracy that is close to the error-free benchmark, demonstrating the robustness of our algorithm under time-varying Rician fading channels.  
}
\section{Conclusions}\label{sec7}
In this paper, we studied the FL design problem in a RIS-assisted communication system. We derived an upper bound on the FL performance by characterizing the performance loss due to device selection and communication noise. We then formulated a unified communication-learning optimization problem with respect to device selection, receiver beamforming, and RIS phase shifts. We proposed a novel algorithm to solve the communication-learning optimization algorithm based on Gibbs sampling and SCA. Finally, we used extensive numerical results to demonstrate the accuracy and convergence improvements of the proposed algorithm compared with the existing approaches. 
The theoretical analysis and numerical experiments in this paper verify the communication-learning tradeoff in the over-the-air FL. The communication system design should be treated together with the learning device selection under a unified framework. This paper provides an initial attempt to characterize the over-the-air  FL performance by taking into account both the communication error and the device selection loss. Further explorations on more general setups such as with non-\iid data distributions or with gradient compression/quantization are promising future research directions.
\appendices
\section{Proof of Lemma \ref{lemmaa}}
\label{appa0}
Plugging \eqref{eq13} into the expression of $\ev_{2,t}$, we have
\begin{align}\label{eqapp2}
		\E[\norm{\ev_{2,t}}_2^2] &{=}\frac{1}{\left(\sum_{m\in \mathcal{M}}K_m\right)^2}\sum_{d=1}^D\E\left[\Big|\sum_{m\in\mathcal{M}} \left( K_m-\frac{\fv^H\hv_m(\thetav)p_m}{\sqrt{\eta}\nu_m}\right)\left({g_m[d]}-{\bar g_m} \right)\Big|^2 +\frac{\abs{\fv^H\nv[d]}^2}{\eta} \right]\nonumber\\
		&\overset{(a)}\geq\frac{\sum_{d=1}^D\E[|\fv^H\nv[d]|^2]}{\eta\left(	\sum_{m\in \mathcal{M}}K_m\right)^2}=\frac{D\sigma^2_n}{\eta\left(	\sum_{m\in \mathcal{M}}K_m\right)^2}.
\end{align}
The equality in $(a)$ holds if $K_m-\frac{\fv^H\hv_m(\thetav)p_m}{\sqrt{\eta}\nu_m}=0$ for $\forall m \in\mathcal{M}$, which is equivalent to the expression of $p_m$ in \eqref{eq15}. Furthermore, by \eqref{eq15} and \eqref{eq11}, we have
$	\abs{p_{m} }^2=\frac{K_m^2 {\eta} \nu_m^2}{\abs{\fv^H\hv_m(\thetav)}^2}	\leq P_0, \forall m \in\mathcal{M}$.
This gives
\begin{align}\label{eqapp1}
	\eta\leq \frac{P_0\abs{\fv^H\hv_m(\thetav)}^2}{K_m^2 \nu_m^2},\forall m \in\mathcal{M}\Leftrightarrow 	\eta\leq\min_{m \in\mathcal{M}} \frac{P_0\abs{\fv^H\hv_m(\thetav)}^2}{K_m^2 \nu_m^2}.
\end{align}
{From \eqref{eqapp2}, we see that $\E[\norm{\ev_{2,t}}_2^2]=\frac{D\sigma^2_n}{\eta\left(	\sum_{m\in \mathcal{M}}K_m\right)^2}$ when $K_m-\frac{\fv^H\hv_m(\thetav)p_m}{\sqrt{\eta}\nu_m}=0$. Therefore, the term $\E[\norm{\ev_{2,t}}_2^2]$ is inversely proportional to $\eta$ in this case. To minimize $\E[\norm{\ev_{2,t}}_2^2]$, it suffices to maximize the value of $\eta$ under \eqref{eqapp1}.}
Taking the maximum $\eta$ in \eqref{eqapp1} gives the $\eta$ in \eqref{eq15}. Finally, plugging $\eta$ from \eqref{eq15} into \eqref{eqapp2}, we obtain \eqref{eq16}, which completes the proof.

\section{Proof of Theorem \ref{the1}}
\label{appa}
First, following the first equation in \cite[Section 3.1]{Friedlander2012}, we bound $\norm{\ev_{1,t}}^2$ as
	\begin{align}\label{eqapp3}
	\norm{\ev_{1,t}}_2^2\leq \frac{4}{K^2} \left(K-\sum_{m\in \mathcal{M}}K_m\right)^2\left(\a_1+\a_2 \norm{\nabla F(\wv_{t})}_2^2\right).
\end{align}
Moreover, from Lemma \ref{lemmaa}, we have
\begin{align}\label{temp33}
\E[\norm{\ev_{2,t}}_2^2]&=\frac{D\sigma^2_n }{P_0\left(\sum_{m\in \mathcal{M}}K_m\right)^2} \max_{m\in \mathcal{M}} \frac{K_m^2 \nu_m^2}{\abs{\fv^H\hv_m(\thetav)}^2}.
\end{align}
For $\forall m$,  we have
\begin{align}
	DK_m^2\nu_m^2&=K_m^2 \sum_{d=1}^D\left(g_{m,t}[d]-\frac{1}{D}\sum_{d^\prime=1}^D g_{m,t}[d^\prime]\right)^2= K_m^2\left(\sum_{d=1}^Dg_{m,t}^2[d]-\frac{1}{D}\left(\sum_{d^\prime=1}^D g_{m,t}[d^\prime]\right)^2\right)\nonumber\\
	&\overset{(a)}\leq K_m^2\left(\sum_{d=1}^Dg_{m,t}^2[d]\right)\overset{(b)}=  \norm{\sum_{(\xv_{mk},y_{mk})\in\mathcal{D}_m}\nabla f(\wv_t;\xv_{mk},y_{mk})}_2^2\nonumber\\
	&\overset{(c)}\leq \left( \sum_{(\xv_{mk},y_{mk})\in\mathcal{D}_m} \norm{\nabla f(\wv_t;\xv_{mk},y_{mk})}_2\right) ^2\overset{(d)}\leq \left( \sum_{(\xv_{mk},y_{mk})\in\mathcal{D}_m}\sqrt{\a_1+\a_2 \norm{\nabla F(\wv_{t})}_2^2}\right) ^2\nonumber\\
	&= K_m^2 \left(\a_1+\a_2 \norm{\nabla F(\wv_{t})}_2^2\right),\label{temp02}
\end{align} 

where $(a)$ is because $\left(\sum_{d^\prime=1}^D g_{m,t}[d^\prime]\right)^2/D\geq 0$; $(b)$ is from the definition of $\gv_{m,t}$ in \eqref{eq03}; $(c)$ is from the triangle inequality; and $(d)$ is from \eqref{eq20}.
Substituting \eqref{temp02} into \eqref{temp33}, we have
\begin{align}\label{temp34}
\E[\norm{\ev_{2,t}}_2^2]&\leq \frac{\sigma^2_n}{P_0\left(\sum_{m\in \mathcal{M}}K_i\right)^2}\left(\a_1+\a_2 \norm{\nabla F(\wv_{t})}_2^2\right)\max_{m\in \mathcal{M}} \frac{K_m^2}{\abs{\fv^H\hv_m(\thetav)}^2}.
\end{align}

Combining \eqref{temp04}, \eqref{eqapp3}, and \eqref{temp34}, we obtain
\begin{align}\label{temp05}
\E[F(\wv_{t+1})]&\leq \E[F(\wv_{t})]-\frac{\norm{\nabla F(\wv_{t})}_2^2}{2\omega}\left(1-2\a_2 d(\mathcal{M},\fv,\thetav) \right)+\frac{\a_1}{\omega}d(\mathcal{M},\fv,\thetav),
\end{align}
where $d(\mathcal{M},\fv,\thetav)$ is defined in \eqref{eq24}.

From \cite[eq. (2.4)]{Friedlander2012}, we have
$\norm{\nabla F(\wv_t)}_2^2\geq 2\mu(F(\wv_t)-F(\wv^\star))$.
Subtracting $F(\wv^\star)$ on both sides of  \eqref{temp05}, applying the above inequality and taking the expectations on both sides, we obtain
\begin{align}\label{temp06}
	\E\left[F(\wv_{t+1})-F(\wv^\star)\right]	\leq \E\left[F(\wv_{t})-F(\wv^\star)\right] \Psi(\mathcal{M},\fv,\thetav)	+\frac{\a_1}{\omega}d(\mathcal{M},\fv,\thetav).
\end{align}
Finally, recursively applying \eqref{temp06} for $t+1$ times, we obtain \eqref{eq23}, which completes the proof.
\section{}
\label{appb}
%
From \eqref{eqappc01}, we see that $\min_{(\fv,\thetav)\in\mathcal{S},\kappa\in \Real}\mathcal{L}^{(i)}(\fv,\thetav,\kappa,\zetav)$ is $-\infty$ if $1-\sum_{m\in\mathcal{M}}K_m^2\mu_m\neq 0$; and is $\min_{(\fv,\thetav)\in\mathcal{S}}
\sum_{m\in\mathcal{M}}\zeta_m c_m^{(i)}-2\Re\{\fv^H( \sum_{m\in\mathcal{M}}\zeta_m\av_m^{(i)}) +\thetav^H( \sum_{m\in\mathcal{M}}\zeta_m\bv_m^{(i)})\}$ otherwise.

By the Cauchy–Schwarz inequality, we have, for any $(\fv,\thetav)\in\mathcal{S}$,
\begin{align}
	\sum_{m\in\mathcal{M}}&\zeta_m c_m^{(i)}-2\Re\left\{\fv^H\left( \sum_{m\in\mathcal{M}}\zeta_m\av_m^{(i)}\right) +\thetav^H\left( \sum_{m\in\mathcal{M}}\zeta_m\bv_m^{(i)}\right) \right\}\nonumber
\end{align}
\begin{align}
	&\geq \sum_{m\in\mathcal{M}}\zeta_mc_m^{(i)} -2\big\lVert{\sum_{m\in\mathcal{M}}\zeta_m\av_m^{(i)}}\big\rVert_2-2\big\lVert{\sum_{m\in\mathcal{M}}\zeta_m\bv_m^{(i)}}\big\rVert_1,\label{eqappc03}
\end{align}
where the equality holds if \eqref{eq33} is satisfied.
Substituting \eqref{eqappc03} to $\min_{(\fv,\thetav)\in\mathcal{S},\kappa\in \Real}\mathcal{L}^{(i)}(\fv,\thetav,\kappa,\zetav)$, we have \eqref{eq34}. 

\ifCLASSOPTIONcaptionsoff
\newpage
\fi
\ifhavebib
{
	\bibliographystyle{IEEEtran}
	\bibliography{ref}
}
\else{
}
\fi
\end{document}